\documentclass[twocolumn]{aastex631}

\usepackage{natbib}
\usepackage{graphicx}
\usepackage{multirow}
\usepackage{mathtools}
\usepackage{makecell}
\usepackage{rotating}

\usepackage{hyperref}
\hypersetup{
    colorlinks=true,
    citecolor=blue,
    linkcolor=blue,
    filecolor=magenta,      
    urlcolor=cyan,
}
\newcommand{\msun}{\mbox{M}_{\odot}}

\newcommand{\hii}{H~\textsc{ii}}
\newcommand{\one}{~\textsc{i}}

\newcommand{\ebvgas}{E(B$-$V)$_{\text{gas}}$}

\newcommand{\sigsfr}{$\Sigma_{\text{SFR}}$}

\newcommand{\temop}{$T_e$(O$^+$)}
\newcommand{\temotp}{$T_e$(O$^{2+}$)}

\newcommand{\densp}{$n_{\text{e}}$(S$^+$)}
\newcommand{\oph}{O$^+$/H$^+$}
\newcommand{\otph}{O$^{2+}$/H$^+$}
\newcommand{\nph}{N$^+$/H$^+$}

\newcommand{\sph}{S$^+$/H$^+$}

\newcommand{\n}{~\textsc{i}}
\newcommand{\ii}{~\textsc{ii}}
\newcommand{\iii}{~\textsc{iii}}
\newcommand{\iv}{~\textsc{iv}}

\newcommand{\oiia}{[O~\textsc{ii}]$\lambda\lambda$7322,7332}
\newcommand{\oiiaa}{[O~\textsc{ii}]$\lambda$7322}
\newcommand{\oiiab}{[O~\textsc{ii}]$\lambda$7332}
\newcommand{\oiis}{[O~\textsc{ii}]$\lambda\lambda$3727,3730}

\newcommand{\pyneb}{\texttt{pyneb}}

\newcommand{\cgsflux}{$\mbox{erg s}^{-1}~\mbox{cm}^{-2}$}

\shorttitle{Auroral [O~\textsc{ii}] at High Redshift}
\shortauthors{Sanders et al.}

\begin{document}

%% Paper title
\title{A Preview of JWST Metallicity Studies at Cosmic Noon: The First Detection of Auroral [O~\textsc{ii}] Emission at High Redshift
\footnote{
The data presented herein were obtained at the W. M. Keck Observatory, which is operated as a scientific partnership among the California Institute of Technology, the University of California and the National Aeronautics and Space Administration. The Observatory was made possible by the generous financial support of the W. M. Keck Foundation.
}
}

%% author list
\author[0000-0003-4792-9119]{Ryan L. Sanders}\altaffiliation{NHFP Hubble Fellow}\affiliation{Department of Physics and Astronomy, University of California, Davis, One Shields Ave, Davis, CA 95616, USA}

\email{email: rlsand@ucdavis.edu}

% alice, leonardo, naveen, mariska, michael, dan, mengtao
% alice, leonardo, michael, naveen, mariska, tucker, dan, mengtao

\author[0000-0003-3509-4855]{Alice E. Shapley}\affiliation{Department of Physics \& Astronomy, University of California, Los Angeles, 430 Portola Plaza, Los Angeles, CA 90095, USA}

\author[0000-0003-1249-6392]{Leonardo Clarke}\affiliation{Department of Physics \& Astronomy, University of California, Los Angeles, 430 Portola Plaza, Los Angeles, CA 90095, USA}

\author{Michael W. Topping}\affiliation{Steward Observatory, University of Arizona, 933 N Cherry Avenue, Tucson, AZ 85721, USA}

\author[0000-0001-9687-4973]{Naveen A. Reddy}\affiliation{Department of Physics \& Astronomy, University of California, Riverside, 900 University Avenue, Riverside, CA 92521, USA}

\author[0000-0002-7613-9872]{Mariska Kriek}\affiliation{Leiden Observatory, Leiden University, P.O.\ Box 9513, NL-2300 AA Leiden, The Netherlands}

\author[0000-0001-5860-3419]{Tucker Jones}\affiliation{Department of Physics and Astronomy, University of California, Davis, One Shields Ave, Davis, CA 95616, USA}

\author{Daniel P. Stark}\affiliation{Steward Observatory, University of Arizona, 933 N Cherry Avenue, Tucson, AZ 85721, USA}

\author[0000-0001-5940-338X]{Mengtao Tang}\affiliation{Department of Physics and Astronomy, University College London, Gower Street, London WC1E 6BT, UK}

\begin{abstract}
We present ultra-deep Keck/MOSFIRE rest-optical spectra of two star-forming galaxies at z=2.18 in the
 COSMOS field with bright emission lines, representing more than 20~hours of total integration.
The fidelity of these spectra enabled the detection of more than 20 unique emission lines for each galaxy,
 including the first detection of the auroral \oiia\ lines at high redshift.
We use these measurements to calculate the electron temperature in the low-ionization O$^+$ zone of the ionized ISM
 and derive abundance ratios of O/H, N/H, and N/O using the direct method.
The N/O and $\alpha$/Fe abundance patterns of these galaxies are consistent with rapid formation timescales
 and ongoing strong starbursts, in accord with their high specific star-formation rates.
These results demonstrate the feasibility of using auroral [O\ii] measurements for accurate metallicity studies at
 high redshift in a higher metallicity regime previously unexplored with the direct method in distant galaxies.
These results also highlight the difficulty in obtaining the measurements required for direct-method metallicities
 from the ground.
We emphasize the advantages that the JWST/NIRSpec instrument will bring to high-redshift metallicity studies,
 where the combination of increased sensitivity and uninterrupted wavelength coverage will yield more than an
 order of magnitude increase in efficiency for multiplexed auroral-line surveys relative to current ground-based facilities.
Consequently, the advent of JWST promises to be the beginning of a new era of precision chemical abundance studies
 of the early universe at a level of detail rivaling that of local galaxy studies.
\end{abstract}

%\keywords{galaxies: ISM --- galaxies: abundances}

%%%%%%%%%%%%%%%%%%%%%%%%%%%%%%%%%%%%%%%%%%%%%%%%%%

%%%%%%%%%%%%%%%%% BODY OF PAPER %%%%%%%%%%%%%%%%%%

\section{Introduction}\label{sec:intro}

%Intro paragraph -- chemical abundances are important for understanding galaxy formation and evolution/gas-phase metallicity a key basic property important for many processes/etc.

The gas-phase oxygen abundance (O/H) in a galaxy provides an essential indication of its evolutionary state,
specifically the aggregate effects of chemical enrichment from past star formation, and dilution from
inward and outward gas flows. The strong correlations between gas-phase oxygen abundance and other galaxy
properties such as stellar mass, star-formation rate (SFR), and gas content has been comprehensively
demonstrated in the local universe \citep[e.g.,][]{tre04,man10,zah14a}.
The form of these relationships has been shown to provide important constraints on the parameters
of star-formation feedback in galaxies, such as the dependence of mass outflow rate on galaxy
mass and the timescales over which SFRs and metallicities vary relative to their equilibrium
values at a given mass \citep[e.g.,][]{peeples2011,andrews2013,dave2017,torrey2018,torrey2019}.

%introduce direct-method as the "gold standard" abundance measures, how these anchor metallicities for larger samples via construction of strong-line metallicity calibrations

The challenge associated with gas-phase oxygen abundance measurements in galaxies lies with
translating spectroscopic measurements of multiple nebular emission line strengths into estimates of metallicity.
One of the most robust methods for estimating oxygen abundances is the so-called ``direct method," in 
which the electron temperature ($T_e$) is derived from the ratio between weak, upper-level auroral line(s)
and stronger, intermediate-level line(s) (e.g., [O\iii]$\lambda 4364$/[O\iii]$\lambda\lambda4960,5008$),
and the electron density ($n_e$) is inferred from the doublet ratio of features such as [S\ii]$\lambda\lambda 6718, 6733$.
These
ionized-gas properties are then used to translate the strengths of strong oxygen lines relative to hydrogen Balmer lines into
the abundance of oxygen ions relative to hydrogen nuclei \citep[e.g.,][]{pem17}.
The direct method has been applied to hundreds of individual \hii\ regions in the Milky Way and nearby galaxies
\citep[e.g.,][]{bresolin2009,ber20}, as well as the integrated spectra of nearby star-forming galaxies
\citep{izo06}. The direct method has also been used to estimate average oxygen abundances
in stacked spectra of large sample of galaxies drawn from the Sloan Digital Sky Survey \citep[SDSS;][]{andrews2013,cur17}.
With such measurements of direct metallicities across the entire population of local
star-forming galaxies, it is possible to calibrate O/H as a function of the ratio of strong emission 
lines (e.g., [N\ii]$\lambda6585$/H$\alpha$, R23=([O\iii]$\lambda\lambda4960,5008$+[O\ii]$\lambda\lambda3727,3730$)/H$\beta$).
%Such calibrations are essential for metallicity studies of the early universe, as the strong lines are far more practical
% to measure at $z>1$ than the faint auroral lines. 
% in terms of oxygen abundances. 
%-general lack of direct metallicities at high redshift to date, small sample assembled from literature in past work (e.g., Sanders+2020, Patricio+2016, I can fill in the "full list" of past high-z auroral detection refs) all based on [O\iii]4364 or O\iii]1661,1666 which naturally limits detected sources to high-ionization objects in which O2+ is dominant form of O, prevents extension to higher metallicities where low-ionization O+ is important. Emphasize that these have basically all been from [O\iii] 4364. 

Since, at high redshift (i.e., $z>1$), these strong-line ratios are much easier to measure than those
involving the faint auroral lines, a calibration between strong-line ratio and oxygen abundance
is essential for metallicity studies of the early universe.
While direct-method metallicity calibrations from the local universe are commonly applied to 
interpret nebular emission-line ratios in high-redshift galaxies, they may
in fact yield biased results for the inferred oxygen abundances. Such biases may occur due
to the evolving physical conditions in the ionized interstellar medium (ISM) of high-redshift galaxies
\citep[e.g.,][]{ste14,sha15,sha19,san20}. It would therefore be ideal
to construct direct-method calibrations of strong emission-line ratios for star-forming
galaxies based on  direct-method oxygen abundances that are also measured at high redshift.
As reviewed by \citet{san20}, a small sample of $\sim 20$ direct metallicity measurements
exists at $z=1.7-3.6$. However, all of these measurements are based on auroral [O\iii] features, either
[O\iii]$\lambda4364$ in the rest-optical, or O\iii]$\lambda\lambda1661,1666$ in the rest-UV.

Given that rest-optical and rest-UV auroral [O\iii] emission is easiest to detect in relatively
metal-poor, high-excitation galaxies, the $z>1$ sample for which such measurements have been
performed is not representative of the full star-forming galaxy population over a wide range
in stellar mass, SFR, and metallicity. Furthermore, while it is possible to trace the relationship
between strong-line ratios and metal abundance over a wide range of metallicity in the local
universe \citep[e.g.,][]{cur17,bia18}, the small dynamic range of galaxy properties probed with auroral
[O\iii] lines at high redshift means that similarly representative correlations for distant galaxies
do not yet exist.

%-as a result, whether strong-line ratio calibrations behave similarly or differently at z>1 is unclear, implications for interpretation of emission lines from large high-z samples

%-something about the advent of JWST and the new era of deep rest-optical spectroscopy we are beginning (not exactly sure how/where to tie it in)

In order to extend  strong-line metallicity calibrations at $z\sim 2-3$ towards higher (i.e., solar) metallicity, we
now require the detection of low-ionization auroral emission lines. Such emission can be detected in the low-excitation
portions of cooler, more metal-rich, star-forming regions. The brightest of 
these features is [O\ii]$\lambda\lambda 7322,7332$. Both \citet{andrews2013} and \citet{cur17} use [O\ii]$\lambda\lambda 7322,7332$
feature to constrain the electron temperature in $z\sim 0$ SDSS galaxies extending up to solar metallicity. However no such
measurements of [O\ii]$\lambda\lambda 7322,7332$ exist in the literature above $z=1$.

In this paper, we present the {\it first}
[O\ii]$\lambda\lambda 7322,7332$ auroral line measurements outside the low-redshift universe. We have detected these
features for two star-forming galaxies at $z=2.18$, located in the COSMOS field. These galaxies were first observed
with the MOSFIRE spectrograph \citet{mcl12} on the Keck~I telescope as part of the MOSFIRE Deep Evolution
Field (MOSDEF) survey \citep{kri15}.
These targets were selected for deep follow-up Keck/MOSFIRE spectroscopy on the basis of their strong rest-frame optical
 emission lines that suggested they would have relatively strong \oiia, which we indeed confirm in this paper.
%Their rest-frame optical strong-line emission properties suggested that they would
%comprise promising targets for follow-up observations of [O\ii]$\lambda\lambda 7322,7332$, which was indeed confirmed
%using Keck/MOSFIRE. 
Measurements such as these will be essential as we attempt to construct direct-method
abundance calibrations for high-redshift rest-frame optical strong-line measurements collected with the {\it James Webb
Space Telescope} ({\it JWST}).

In \S\ref{sec:obs}, we describe our new MOSFIRE observations and emission-line measurements.
In \S\ref{sec:abun}, we present the physical conditions and
chemical abundances of the two galaxies studied here. 
In \S\ref{sec:discussion}, we discuss the implications of our results for chemical abundance
calibrations at high redshift; the prospects for conducting direct metallicity studies with
the new capabilities of {\it JWST}; the constraints obtained for the chemical abundance patterns
among oxygen, nitrogen, and iron; the (lack of) significant active galactic nucleus (AGN) activity
in our target galaxies; and, finally, the detection of broad rest-frame optical emission, signaling the
presence of galaxy-scale outflows. In \S\ref{sec:summary}, we summarize our key results
and conclusions.  Throughout, we adopt cosmological parameters of
$H_0=70\mbox{ km  s}^{-1}\mbox{ Mpc}^{-1}$, $\Omega_m=0.30$, and
$\Omega_{\Lambda}=0.7$, and a \citet{cha03} initial mass function (IMF).
Rest-frame wavelengths of emission lines are given in the vacuum.
We adopt solar abundance values of 12+log(O/H)$_{\odot}$=8.69, 12+log(N/H)$_{\odot}$=7.83, log(N/O)$_{\odot}$=$-0.86$,
 and log(Fe/H)$_{\odot}$=$7.50$ as number density fractions,
 and the bulk metallicity by mass fraction $Z_\odot=0.014$ \citep{asp09}.

\section{Observations and Measurements}\label{sec:obs}

\subsection{Targets, observations, and data reduction}

Detecting faint temperature-sensitive auroral emission lines at high redshifts requires
 deep near-infrared spectroscopy of targets displaying bright line emission.
Accordingly, we obtained ultra-deep observations of targets in the COSMOS CANDELS \citep{gro11,koe11}
 field using the MOSFIRE instrument \citep{mcl12} on the 10~m Keck~I telescope.
The primary targets were selected from the MOSDEF survey \citep{kri15}, which obtained
 rest-optical spectroscopy of $\sim1500$ galaxies at $z=1.4-3.8$ using MOSFIRE,
 with a typical depth of 2~hours per filter.
Fluxes and ratios of strong emission lines measured in the MOSDEF spectra were used to
 predict the flux of auroral [O\iii]$\lambda$4364 and [O\ii]$\lambda\lambda$7322,7332.
A slitmask was then designed that maximized the number of targets with predicted auroral
 line fluxes that could be detected in feasible integration times with MOSFIRE.
The final mask in COSMOS included 5 galaxies at $z=2.0-3.5$ targeted for [O\iii]$\lambda$4364 that will be
 presented in a future work
 and two galaxies at $z=2.18$ targeted for \oiia\ that are the subject of the current analysis.
The relatively high excitation level (typically associated with larger $T_e$)
 and strong reddening-corrected \oiis\ flux displayed in the
 MOSDEF spectra of the latter targets suggested bright auroral \oiia\ lines that fall
 in the $K$-band filter at this redshift.
These two galaxies are identified by their ID numbers 19985 and 20062 in the v4.1 photometric
 catalogs of the 3D-HST survey \citep{ske14,mom16}.
Their coordinates are given in Table~\ref{tab:properties}.
These targets are among the brightest emission-line galaxies in the MOSDEF survey data set
 based on their observed H$\alpha$ and [O\iii]$\lambda$5008 fluxes.

\begin{table}
 \centering
 \caption{Target properties.
 }\label{tab:properties}
 \begin{tabular}{ l l l }
\hline\hline
ID  &  19985  &  20062  \\
   \hline
R.A. (J2000)  & 10:00:14.484 & +02:22:57.98 \\
Dec. (J2000)  & 10:00:16.436 & +02:23:00.79 \\
$z$  & 2.18796 & 2.18541  \\
log($M_*$/M$_{\odot}$)  & 9.92$\pm$0.03 & 10.21$\pm$0.07 \\
log($t_{\text{age}}$/yr)\tablenotemark{a}  & 7.50$\pm$0.05 & 8.2$\pm$0.20 \\
log($\tau$/yr)\tablenotemark{a}  & 10.0$\pm$1.0 & 8.2$\pm$1.0 \\
$A_{V,\text{stars}}$  & 1.4$\pm$0.02 & 1.2$\pm$0.03 \\
SFR (M$_{\odot}$ yr$^{-1}$)  & 208$\pm$8 & 265$\pm$8 \\
sSFR (Gyr$^{-1}$)  & 25.0$\pm$2.0 & 16.3$\pm$2.8 \\
$R_{\text{eff}}$ (kpc)\tablenotemark{b}  & 1.34 & 1.46 \\
$\Sigma_{\text{SFR}}$\tablenotemark{c} (M$_{\odot}$ yr$^{-1}$ kpc$^{-2}$)  & 18.5$\pm$0.7 & 19.8$\pm$0.6 \\
\hline
 \end{tabular}
\tablenotetext{a}{Stellar population age and timescale in a delayed-$\tau$ star-formation history of the form $\text{SFR}\propto t_{\text{age}} e^{-t_{\text{age}}/ \tau}$.}
\tablenotetext{b}{Half-light elliptical semi-major axis from \citet{van14}.}
\tablenotetext{c}{\sigsfr=SFR/2$\pi$$R_{\text{eff}}^2$}
\end{table}

The MOSFIRE mask was observed for a total of 15.4~hours on 5 nights spanning
 13 January 2019 to 4 March 2021, with integrations of
 1.9~h in $J$ band, 8.0~h in $H$ band, and 5.5~h in $K$ band.
The median seeing, measured from the spatial profile of a star on the mask,
 was 0.79\arcsec, 0.51\arcsec, and 0.44\arcsec\ in $J$, $H$, and $K$, respectively.
We adopted the standard observing strategy used in the MOSDEF survey, specifically
 an ABA$^\prime$B$^\prime$ dither pattern with an inner/outer nod of 1.2\arcsec/1.5\arcsec,
 2 minute exposures in the $J$ and $H$ bands, and 3 minute exposures in the $K$ band.
Slit widths were 0.7\arcsec, yielding spectral resolutions of $R\sim3300$, 3650, and 3600 in $J$, $H$, and $K$, respectively.
The data were reduced using a custom IDL pipeline designed for the MOSDEF survey
 and described in \citet{kri15}, the product of which is fully calibrated two-dimensional
 science spectra for each slit on the mask.
The program \texttt{bmep}\footnote{https://github.com/billfreeman44/bmep} \citep{fre19}
 was used to obtain one-dimensional science and error spectra using
 an optimal extraction \citep{hor86}.
%A slit loss correction factor was then calculated using the seeing,
% \textit{HST} WFC3/F160W imaging, and slit geometry and applied to each 1-D science spectrum \citep{red15}.

In the MOSDEF reduction pipeline, the absolute flux calibration is achieved via a star placed
 on one of the MOSFIRE slits.
In each filter, the measured continuum spectrum for the slit star is scaled to match
 the cataloged photometry of the star, and the scaling factor is then applied to all targets on a mask.
The slit star on our target mask unfortunately dithered directly on top of a galaxy of comparable
 brightness separated by 2.7\arcsec.
As a result, the stellar spectrum was oversubtracted and the flux calibration of the mask was biased.
However, the targets of this analysis were observed on a MOSDEF survey mask that has an isolated
 slit star such that the MOSDEF flux calibration is reliable.
To achieve a robust flux calibration of the new observations, for each target, we measured the fluxes
 of the brightest emission line in each filter (\oiis\ in $J$, [O\iii]$\lambda$5008 in $H$, and H$\alpha$ in $K$)
 in both sets of spectra, noting that each line is detected at $>25\sigma$.
In each filter, we then scaled the new spectrum by a multiplicative factor such that the line flux matched
 that measured from the MOSDEF spectrum,
 the latter of which has been corrected for slit losses as described in \citet{red15} and \citet{kri15}.

\begin{figure*}
 \includegraphics[width=\textwidth]{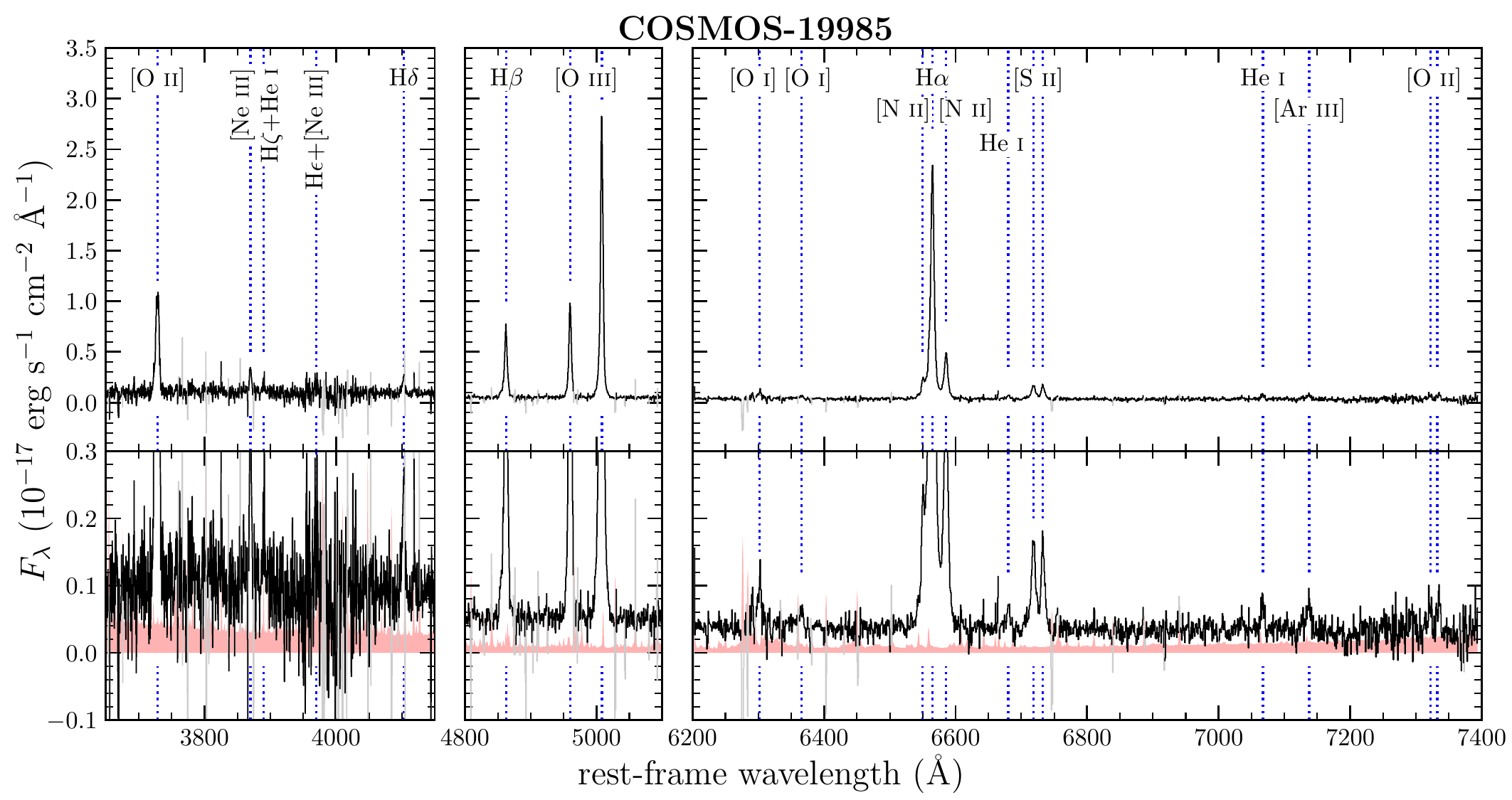}
 \includegraphics[width=\textwidth]{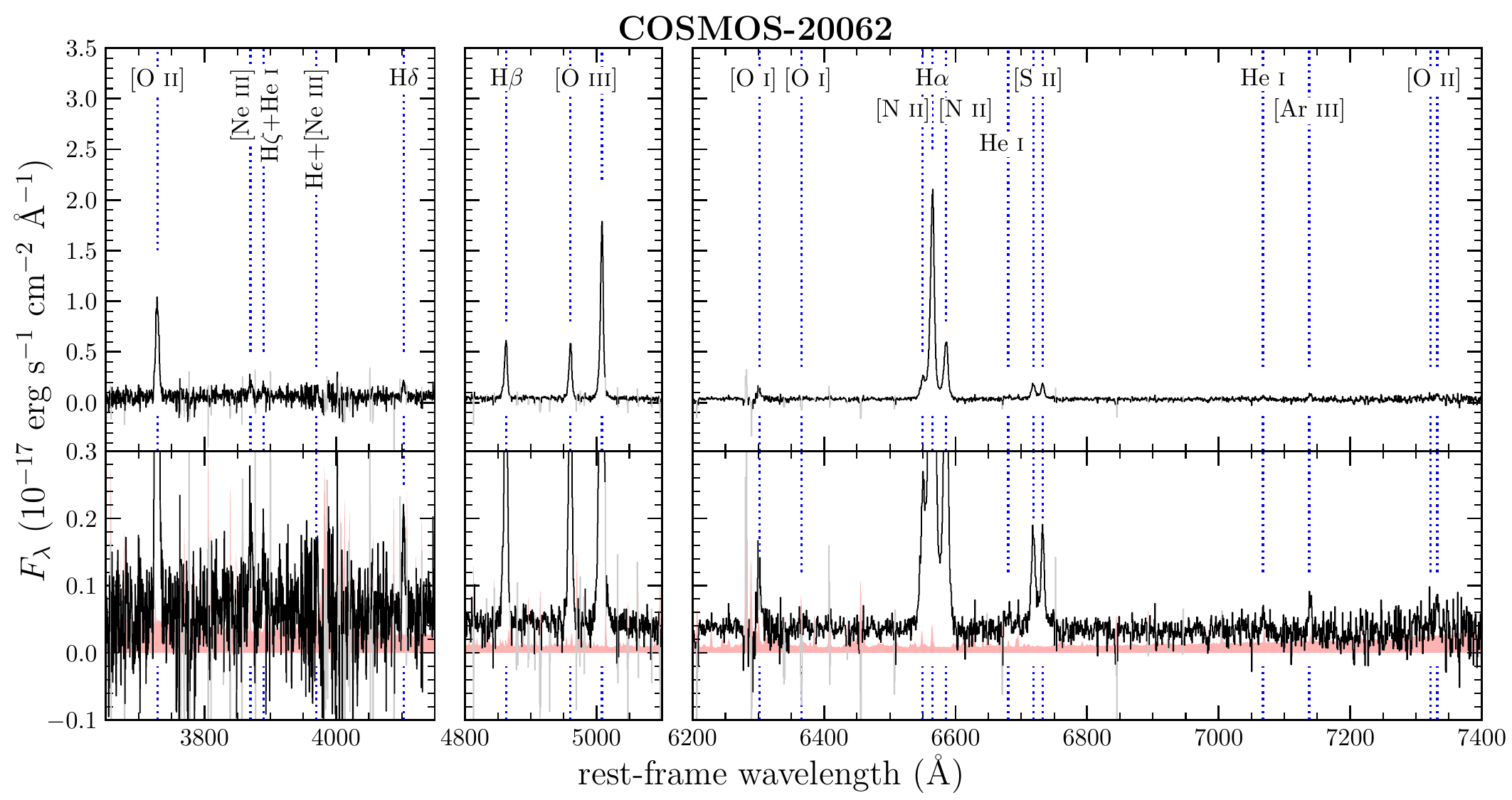}
 \centering
 \caption{
Near-infrared spectra of COSMOS 19985 (top) and 20062 (bottom), showing spectral features as a function of rest-frame wavelength
 observed in the $J$ (left), $H$ (middle), and $K$ (right) bands.
Regions affected by strong sky lines are grayed out.
The shaded light red region shows the $1\sigma$ error spectrum.
For each target, the upper panels present the full dynamic range of the spectrum, while the lower panels
 show the same spectrum zoomed in to highlight weak emission lines.
Detected emission lines are labeled according to the ionic species.
Blue dashed lines indicate the vacuum wavelengths of each transition.
}\label{fig:spectra}
\end{figure*}

To maximize the signal-to-noise ratio, we combined the MOSDEF spectra with our new observations.
The final 1-D science spectra were constructed by taking the inverse-variance weighted mean of the flux density
 at each wavelength pixel.
We note that the MOSDEF $K$-band spectra only extend to 2.315~$\mu$m ($\approx$7270~\AA\ rest-frame), such that
 the reddest part of the $K$-band spectrum covering \oiia\ includes only the new spectroscopy.
All other measured spectral features are covered by both data sets.
The final spectra thus have effective total integration times of 3.8~h, 10.0~h, and 7.5~h in $J$, $H$, and $K$,
 respectively, reaching typical 3$\sigma$ limiting line fluxes of $7\times10^{-18}$, $3\times10^{-18}$,
 and $4\times10^{-18}$~$\mbox{erg s}^{-1}~\mbox{cm}^{-2}$ in spectral regions free of strong sky lines.
These spectra, in which a number of strong and weak emission lines are visible, are presented in Figure~\ref{fig:spectra}.

\subsection{Spectral energy distribution (SED) fitting}

Galaxy properties were inferred from fitting stellar population models to photometry measured in 43 filters
 spanning rest-frame 1200~\AA\ to 2.5~$\mu$m, as cataloged by the 3D-HST survey team \citep{ske14,mom16}.
We used the SED-fitting code FAST \citep{kri09} in combination with the flexible stellar population synthesis models
 of \citet{con10}.
A delayed-$\tau$ star-formation history, solar metallicity, a \citet{cal00} dust attenuation curve, and a \citet{cha03} IMF were assumed.
Before fitting, the observed near-infrared photometric measurements were corrected for the contribution from strong emission
 lines using initial line flux estimates following \citet{san21}.
Due to the large emission-line equivalent widths in our targets, this correction is important, with emission lines accounting
 for approximately 20\%, 40\%, and 60\% of the photometric flux density in the $J$-, $H$-, and $K$-band filters.
Derived stellar masses are $\sim0.5$~dex higher without the emission-line correction.
This process yielded estimates of stellar mass, stellar continuum reddening, stellar population age, and a best-fit
 model of the stellar continuum.
In the $H$ and $K$ bands, where the continuum is detected significantly (S/N per pixel $\approx$5),
 the best-fit SED model agrees with the observed MOSFIRE spectrum to better than 15\%,
 which validates our absolute and band-to-band flux calibration.
The best-fit stellar population parameters are given in Table~\ref{tab:properties}.

\subsection{Emission line measurements}

Emission-line fluxes were measured by fitting Gaussian line profiles to the 1D science spectra.
The continuum under the lines was taken to be the best-fit stellar continuum model from SED fitting.
Similar results are obtained if we instead fit a constant or linear local continuum under each line.
Using the best-fit SED model has the advantage of self-consistently accounting for absorption under hydrogen Balmer lines.
All emission features were fit with single Gaussian profiles except for 
\oiis, [S\ii]$\lambda\lambda$6718,6733, and \oiia\ that were simultaneously fit with a double Gaussian, and
H$\alpha$ and [N\ii]$\lambda\lambda$6550,6585 that were simultaneously fit with a triple Gaussian.
%All other emission lines are fit with a single Gaussian profile.
The systemic redshifts were taken from the average redshift of H$\alpha$ and [O\iii]$\lambda$5008,
 and are consistent with the previously-published MOSDEF redshifts.
The centroids and velocity widths of weak lines were set by the redshift and velocity width measured for
 H$\alpha$ and [O\iii]$\lambda$5008.
Uncertainties on line fluxes were estimated by perturbing the spectrum according to the error spectrum
 and remeasuring the line fluxes 500 times, where the $1\sigma$ uncertainties were taken to be half of the
 16th-to-84th percentile width of the resulting flux distributions.
Emission-line fluxes and uncertainties are presented in Table~\ref{tab:lines}.
More than 20 unique emission lines are detected at $\ge3\sigma$ significance for each target.

\begin{table}
 \centering
 \caption{Observed line fluxes and reddening-corrected line ratios. For non-detected lines, 3$\sigma$ upper limits are given.
 }\label{tab:lines}
 \begin{tabular}{ l l l }
\hline\hline
\multicolumn{3}{c}{$F_{obs}(\lambda$) ($10^{-17}~\mbox{erg s}^{-1}~\mbox{cm}^{-2}$)}\\
\hline
Line  &  19985  &  20062  \\
   \hline
$[$O\ii$]~$$\lambda$3727,3730$^\dag$  &  $20.93$$\pm$$0.45$  &  $19.51$$\pm$$1.42$  \\
$[$Ne\iii$]$~$\lambda$3870  &  $3.49$$\pm$$0.23$  &  $2.30$$\pm$$0.22$  \\
H$\zeta$+He\n~$\lambda$3890$^\dag$  &  $2.29$$\pm$$0.25$  &  $1.70$$\pm$$0.20$  \\
H$\zeta$~$\lambda$3890\tablenotemark{a}  &  $1.28$$\pm$$0.29$  &  $1.00$$\pm$$0.25$  \\
H$\epsilon$+$[$Ne\iii$]$~$\lambda$3970$^\dag$  &  $2.33$$\pm$$0.37$  &  $1.61$$\pm$$0.31$  \\
H$\epsilon$~$\lambda$3971\tablenotemark{b}  &  $1.28$$\pm$$0.38$  &  $0.92$$\pm$$0.32$  \\
$[$S\ii$]$~$\lambda$4078  &  $<$$0.48$  &  $<$$0.71$  \\
H$\delta$~$\lambda$4103  &  $3.03$$\pm$$0.24$  &  $2.46$$\pm$$0.18$  \\
He\ii~$\lambda$4686  &  $<$$0.25$  &  $<$$0.39$  \\
H$\beta$~$\lambda$4863  &  $11.60$$\pm$$0.20$  &  $10.10$$\pm$$0.12$  \\
$[$O\iii$]$~$\lambda$4960  &  $13.90$$\pm$$0.14$  &  $9.02$$\pm$$0.12$  \\
$[$O\iii$]$~$\lambda$5008  &  $45.40$$\pm$$0.12$  &  $29.20$$\pm$$0.08$  \\
$[$O\n$]$~$\lambda$6302  &  $1.46$$\pm$$0.15$  &  $1.62$$\pm$$0.22$  \\
$[$S\iii$]$~$\lambda$6314  &  $<$$0.81$  &  $<$$0.81$  \\
$[$O\n$]$~$\lambda$6366  &  $0.46$$\pm$$0.10$  &  $<$$0.56$  \\
$[$N\ii$]$~$\lambda$6550  &  $6.26$$\pm$$0.10$  &  $7.49$$\pm$$0.15$  \\
H$\alpha$~$\lambda$6565  &  $48.80$$\pm$$0.13$  &  $47.60$$\pm$$0.14$  \\
$[$N\ii$]$~$\lambda$6585  &  $11.00$$\pm$$0.07$  &  $15.00$$\pm$$0.11$  \\
He\n~$\lambda$6680  &  $0.62$$\pm$$0.07$  &  $0.69$$\pm$$0.10$  \\
$[$S\ii$]$~$\lambda$6718  &  $3.33$$\pm$$0.10$  &  $3.82$$\pm$$0.15$  \\
$[$S\ii$]$~$\lambda$6733  &  $2.74$$\pm$$0.09$  &  $3.23$$\pm$$0.14$  \\
He\n~$\lambda$7067  &  $0.86$$\pm$$0.12$  &  $0.46$$\pm$$0.14$  \\
$[$Ar\iii$]$~$\lambda$7138  &  $0.88$$\pm$$0.15$  &  $0.84$$\pm$$0.17$  \\
$[$O\ii$]$~$\lambda$7322  &  $0.74$$\pm$$0.19$  &  $<$$0.67$  \\
$[$O\ii$]$~$\lambda$7332  &  $0.75$$\pm$$0.20$  &  $0.89$$\pm$$0.24$  \\
\hline
% \end{tabular}
% \begin{tabular}{ l l }
\multicolumn{3}{c}{Reddening-Corrected Line Ratios} \\
\hline
\ebvgas  &  $0.40$$\pm$$0.02$  &  $0.52$$\pm$$0.01$  \\
log([O\iii]$\lambda$5008/H$\beta$)  &  $0.57$$\pm$$0.01$  &  $0.43$$\pm$$0.01$  \\
log([O\ii]$\lambda$3728/H$\beta$)  &  $0.44$$\pm$$0.01$  &  $0.53$$\pm$$0.03$  \\
log(R23)\tablenotemark{c}  &  $0.89$$\pm$$0.01$  &  $0.84$$\pm$$0.02$  \\
log([O\iii]$\lambda$5008/[O\ii]$\lambda$3728)  &  $0.13$$\pm$$0.01$  &  $-0.09$$\pm$$0.03$  \\
log([Ne\iii]$\lambda$3870/[O\ii]$\lambda$3728)  &  $-0.80$$\pm$$0.03$  &  $-0.95$$\pm$$0.05$  \\
log([N\ii]$\lambda$6585/H$\alpha$)  &  $-0.65$$\pm$$0.01$  &  $-0.50$$\pm$$0.01$  \\
log(O3N2)\tablenotemark{d}  &  $1.22$$\pm$$0.01$  &  $0.94$$\pm$$0.01$  \\
log([S\ii]$\lambda\lambda$6718,6733/H$\alpha$)  &  $-0.92$$\pm$$0.01$  &  $-0.84$$\pm$$0.01$  \\
log([O\n]$\lambda$6302/H$\alpha$)  &  $-1.50$$\pm$$0.05$  &  $-1.44$$\pm$$0.06$  \\
$\frac{[\mbox{O\ii}]\lambda7322,7332}{[\mbox{O\ii}]\lambda3727,3730}$  &  $0.027$$\pm$$0.005$  &  $0.017$$\pm$$0.005$  \\
\hline
 \end{tabular}
%\tablenotetext{a}{Sum of the [O\ii]$\lambda$3727 and [O\ii]$\lambda$3730 lines.}
\\$^\dag$ Blended lines.
\tablenotetext{a}{Derived by removing the blended He\n~$\lambda$3890 flux estimated using He\n~$\lambda$6680 and He\n~$\lambda$7067.}
\tablenotetext{b}{Derived by removing the blended [Ne\iii]$\lambda$3969 flux estimated using [Ne\iii]$\lambda$3870.}
\tablenotetext{c}{R23=([O\iii]$\lambda\lambda$4960,5008+[O\ii]$\lambda\lambda$3727,3730)/H$\beta$.}
\tablenotetext{d}{O3N2=([O\iii]$\lambda$5008/H$\beta$)/([N\ii]$\lambda$6585/H$\alpha$).}
\end{table}

A single line flux is reported for blended features.
While the \oiis\ doublet centroids are resolved at the MOSFIRE spectral resolution ($R>3000$),
 the relatively broad line widths (FWHM$\sim$300~km~s$^{-1}$) prevent cleanly separating the doublet
 components.
As such, we report only the total \oiis\ flux and refer to this sum as [O\ii]$\lambda$3728.
H$\epsilon$ is blended with [Ne\iii]$\lambda$3969.
We derived the H$\epsilon$ flux by subtracting the blended [Ne\iii]$\lambda$3969 flux, where
 the latter was inferred from [Ne\iii]$\lambda$3870, leveraging the fixed ratio of the two
 lines of [Ne\iii]$\lambda$3870/$\lambda$3969=3.32 calculated with \pyneb\ \citep{lur13,lur15}.
We estimated the He\n~$\lambda$3890 flux from the detected He\n~$\lambda\lambda$6680,7067 lines
 using \pyneb, assuming the electron temperature and density derived below,
 and used this flux to infer the deblended H$\zeta$ flux.

\subsection{Reddening correction, line ratios, and SFR}

Nebular reddening, \ebvgas, was derived assuming a \citet{car89} Milky Way extinction curve with $R_V=3.1$
 via the Balmer decrement H$\alpha$/H$\beta$, where the final intrinsic ratio
 used (2.83 for 19985 and 2.89 for 20062) was calculated with \pyneb\ assuming the electron temperature
 and density calculated below.
Line fluxes were corrected for reddening using \ebvgas\ and the \citet{car89} extinction curve.
\ebvgas\ derived from H$\delta$/H$\alpha$, H$\epsilon$/H$\alpha$, and H$\zeta$/H$\alpha$ agree
 with our fiducial value within 2$\sigma$, suggesting that the adopted dust curve is
 appropriate for these targets and the reddening correction is robust down to blue
 wavelengths near [O\ii]$\lambda$3728 (see also \citealt{red20}).
The reddening-corrected line ratios are presented in Table~\ref{tab:lines}.
SFR was calculated using the dust-corrected H$\alpha$ luminosity
 assuming the conversion factor of \citet{hao11} adjusted to a \citet{cha03} IMF.

\subsection{Atomic data}

When calculating temperatures, densities, and chemical abundances below with \pyneb, we adopted
 the atomic data recommended by \citet{ber15} as follows.
For [O\ii] and [O\iii], we used the collision strengths from \citet{kis09} and \citet{sto14}, respectively.
The collision strengths of \citet{tay10} were used for [S\ii], while those of \citet{tay11} were used for [N\ii].
The radiative transition probabilities were taken from \citet{fro04} for all ions.
When using other atomic data sets available in \pyneb, we found that the derived temperatures, densities,
 and abundances changed by $\le0.1$~dex relative to our fiducial set, smaller than the derived uncertainties on these properties.
Systematic uncertainties associated with atomic data are thus not a major contribution to the error budget in this analysis.

\section{Physical Conditions and Chemical Abundances}\label{sec:abun}

\subsection{Auroral \oiia\ lines}

\begin{figure}
 \includegraphics[width=\columnwidth]{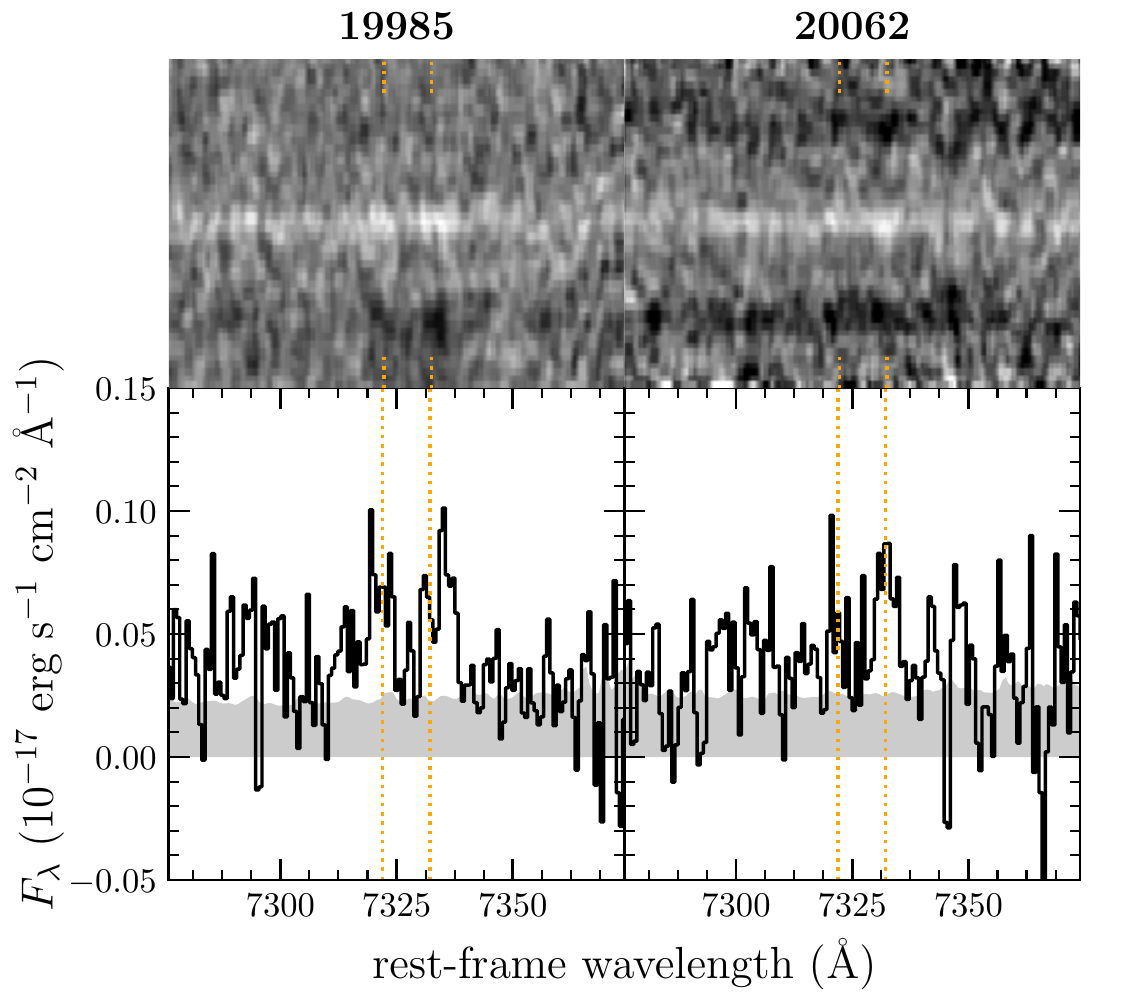}
 \centering
 \caption{
Spectral region around the auroral \oiia\ emission lines for 19985 (left) and 20062 (right).
The bottom panels show the 1D science spectra, with the error spectrum displayed in gray.
The vacuum centroid wavelengths of the transitions are marked in orange.
Both lines are detected at $>3\sigma$ for 19985, while only [O\ii]$\lambda$7332 is formally detected for 20062.
The 2D spectrum (top) confirms these detections.
}\label{fig:oiia}
\end{figure}

Auroral [O\ii] emission lines are detected in the spectra of both targets,
 the first time these temperature-sensitive lines have been detected beyond the low-redshift universe.
Figure~\ref{fig:oiia} shows the auroral [O\ii] lines in the 1D and 2D science spectra.
Both components are detected at $>3\sigma$ in the spectrum of 19985, with significances of
 3.9$\sigma$ and 3.8$\sigma$ for \oiiaa\ and \oiiab, respectively, and a combined significance of 5.5$\sigma$ for the doublet.
For 20062, \oiiaa\ is formally undetected (1.5$\sigma$), while \oiiab\ is detected at the 3.7$\sigma$ level.
%For 20062, only the redder component is formally detected at a significance of 3.7$\sigma$.

The electron temperature in the O$^+$ zone, \temop, can be calculated from the auroral-to-strong line
 ratio \oiia/\oiis.
Since only \oiiab\ is detected in the spectrum of 20062, we need an estimate of the total
 \oiia\ flux to perform this calculation.
The two auroral [O\ii] lines originate from the same upper level and therefore
 have a fixed flux ratio of \oiiaa/\oiiab=1.17 determined by the ratio of radiative transition probabilities.
While the \oiiaa\ flux can be estimated by multiplying the detected \oiiab\ flux of 20062 by 1.17,
 this method yields a \oiiaa\ flux that is 50\% larger than the 3$\sigma$ upper limit reported in Table~\ref{tab:lines},
 suggesting that the \oiiab\ flux of 20062 may be overestimated.
%When searching for weak lines near the detection threshold, a bias arises due to noise fluctuations in which the line flux
% of detections will tend to be overestimated on average because upward noise fluctuations will push lines above the detection
% threshold while downward fluctuations would remove a line from consideration.
To infer the total \oiia\ flux including information from the \oiiaa\ upper limit, we instead fit a model in which the
 two doublet components are fixed to the theoretical ratio and perform a $\chi^2$ minimization to the
 flux and uncertainty of each line inferred from Gaussian fitting.
Since the doublet ratio is fixed, the only free parameter in this model is the total auroral [O\ii] flux.
20062 has observed Gaussian-fit fluxes of 3.2($\pm$2.2)$\times10^{-18}$
 and 8.9($\pm$2.4)$\times10^{-18}$~\cgsflux\ for \oiiaa\ and \oiiab, respectively.
Fitting with the fixed flux ratio model yields a best-fit total \oiia\ flux of 1.1($\pm$0.3)$\times10^{-17}$~\cgsflux.
The best-fit flux of the individual doublet components from the fixed flux ratio model are within 1.5$\sigma$ of the
 Gaussian-fit line fluxes.
%We use this auroral [O\ii] flux to calculate the \oiia/\oiis\ ratio reported in Table~\ref{tab:lines}.

For consistency, we use this same process to infer the total \oiia\ flux of 19985,
 noting that 19985 has a directly measured doublet ratio of \oiiaa/\oiiab=0.98$\pm$0.38, fully consistent with
 the theoretical ratio of 1.17.
This target has observed Gaussian-fit fluxes of 7.4($\pm$1.9)$\times10^{-18}$
 and 7.5($\pm$2.0)$\times10^{-18}$~\cgsflux\ for \oiiaa\ and \oiiab, respectively, and a summed doublet
 flux of 1.49($\pm$0.28)$\times10^{-17}$~\cgsflux.
The best-fit total flux from the fixed flux ratio model is 1.48($\pm$0.28)$\times10^{-17}$~\cgsflux.
These best-fit total flux values were used in the calculation of the \oiia/\oiis\ ratios
 given in Table~\ref{tab:lines}.

\subsection{Temperatures and Densities}

We use the python package \pyneb\ to derive electron temperatures ($T_e$) and densities ($n_e$),
 presented in Table~\ref{tab:temden}.
Electron density is calculated from the [S\ii] doublet using the ratio [S\ii]$\lambda$6718/[S\ii]$\lambda$6733.\footnote{The \oiis\ doublet
 was not robusty deblended due to the relatively broad line widths in these targets, and thus does not provide reliable $n_e$ constraints.}
We find values of $n_e\approx200$~cm$^{-3}$ for both targets, consistent with the typical density of
 $200-300$~cm$^{-3}$ found for galaxies at $z\sim2-3$ \citep[e.g.,][]{ste14,shim15,san16a,kaa17,str17}
 and elevated relative to $z\sim0$ main-sequence galaxies that typically have $n_e<100$~cm$^{-3}$ \citep{kas19b}.

The temperatures in the low-ionization O$^+$ zone, \temop,
 are calculated using the \oiia/\oiis\ ratios given in Table~\ref{tab:lines}.
We find \temop=12,440$\pm$1,680~K and 9,330$\pm$1,350~K for 19985 and 20062, respectively.
That 19985 has a higher \temop\ than 20062 is in accord with the excitation-sensitive line ratios of these galaxies.
19985 has higher [O\iii]/H$\beta$, [O\iii]/[O\ii], and [Ne\iii]/[O\ii] than 20062 indicating a higher level
 of excitation and ionization that are usually associated with lower metallicity and higher temperature.
These values of \temop, near $10^4$~K, are typical of moderately subsolar-metallicity ($\sim0.2-0.5~Z_\odot$) \hii\ regions
 in the local universe \citep{ber15,ber20,cro15,cro16,rog21}.

\begin{table}
 \centering
 \caption{Electron temperatures and densities, and ionic and total elemental abundances calculated using the direct method.
 }\label{tab:temden}
 \begin{tabular}{ l l l }
   \hline\hline
ID  &  19985  &  20062  \\
   \hline
\temop~~(K)  &  12440$\pm$1680  &  9330$\pm$1350  \\ 
\temotp\tablenotemark{a}~~(K)  &  13480$\pm$2620  &  9040$\pm$2390  \\ 
\densp~~(cm$^{-3}$)  &  200$\pm$60  &  210$\pm$70  \\ 
\hline
12+log(\oph)  &  $7.64$$\pm$$0.22$  &  $8.23$$\pm$$0.29$  \\ 
12+log(\otph)  &  $7.72$$\pm$$0.23$  &  $8.14$$\pm$$0.43$  \\ 
12+log(O/H)  &  $7.98$$\pm$$0.22$  &  $8.49$$\pm$$0.35$  \\ 
12+log(\nph)  &  $6.89$$\pm$$0.15$  &  $7.34$$\pm$$0.23$  \\ 
12+log(N/H)  &  $7.23$$\pm$$0.17$  &  $7.60$$\pm$$0.29$  \\ 
12+log(N/O)  &  $-0.75$$\pm$$0.12$  &  $-0.89$$\pm$$0.19$  \\ 
12+log(\sph)  &  $5.73$$\pm$$0.15$  &  $6.10$$\pm$$0.21$  \\ 
\hline
 \end{tabular}
\tablenotetext{a}{Derived from \temop\ assuming the relation of \citet{cam86}.}% The uncertainty includes a 1300~K intrinsic scatter in \temotp at fixed \temop\ \citep{rog21}.}
\end{table}

For other species in the low-ionization zone, namely N$^+$ and S$^+$, we also use the O$^+$ temperature:
\begin{equation}
T_e(\mbox{N}^+) = T_e(\mbox{S}^+) = T_e(\mbox{O}^+)
\end{equation}
Following \citet{rog21}, we add the intrinsic scatter observed in the relations between these temperatures
 in local \hii\ region samples in quadrature with the uncertainty propagated from \temop,
 adopting an intrinsic scatter of 1,000~K \citep{ber20}.

We cannot directly calculate the temperature in the high-ionization O$^{2+}$ zone, \temotp, because [O\iii]$\lambda$4364 at $z=2.18$
 falls at a wavelength of low atmospheric transmission and beyond the reach of ground-based observatories.
Instead, we adopt the \temop$-$\temotp\ relation of \citet{cam86}:
\begin{equation}\label{eq:t2t3}
T_e(\mbox{O}^+) = 0.7 \times T_e(\mbox{O}^{2+}) + 3000~K
\end{equation}
It has recently been shown that there is considerable scatter in the \temop$-$\temotp\ relation,
 with an intrinsic scatter of $\approx$1,300~K in \temotp\ at fixed \temop\ \citep{ber20,rog21}.
Following \citet{rog21}, we add an uncertainty of 1,300~K in quadrature with the error propagated
 from \temop\ when calculating \temotp\ using equation~\ref{eq:t2t3}.
Partially due to the large intrinsic scatter, there is also
 considerable uncertainty about the shape of the \temop$-$\temotp\ relation.
If we instead adopt a 1:1 relation, the derived O/H changes by only $\approx0.1$~dex.

\subsection{Ionic and Total Abundances}

Ionic abundances are calculated using \pyneb\ assuming the temperatures appropriate to each ion as described above.
We assume that the density is constant throughout the nebula, adopting \densp\ for all ions.
%In practice, the effects of density on abundance calculations are negligible at the typical densities observed
% in \hii\ regions ($<1,000$~cm$^{-3}$).
The derived ionic and total abundances are presented in Table~\ref{tab:temden}.

For oxygen, we calculate \oph\ using dust-corrected [O\ii]$\lambda$3728/H$\beta$ and \temop,
 and \otph\ using [O\iii]$\lambda$5008/H$\beta$ and \temotp\ estimated
 with equation~\ref{eq:t2t3}.
We take the total oxygen abundance to be the sum of these two phases:
\begin{equation}
\frac{\text{O}}{\text{H}} = \frac{\text{O}^+}{\text{H}^+} + \frac{\text{O}^{2+}}{\text{H}^+}
\end{equation}
With an ionization energy of 54.9~eV,
 O$^{3+}$ is found to be only a small ($\lesssim$5\%) contribution even in extremely high-ionization sources \citep{ber18,ber21}
 and can safely be ignored.
We find the direct-method oxygen abundances of 19985 and 20062 to be 12+log(O/H)=7.98$\pm$0.22 (0.2$\pm$0.1~$Z_\odot$)
 and 12+log(O/H)=8.49$\pm$0.35 (0.6$^{+0.7}_{-0.3}$~$Z_\odot$), respectively.

Since we have directly constrained \temop, we can calculate direct-method abundances for the low-ionization species
 N$^+$ and S$^+$.
\nph\ is derived using [N\ii]$\lambda$6585/H$\alpha$ and \temop.
To calculate the total N/H, an ionization correction is required to account for N$^{2+}$ for which no associated lines are observed.
We adopt the commonly-used correction factor that leverages the similar ionization potential energies of O and N \citep{pei67}:
\begin{equation}
\text{ICF(N)} = \frac{\text{N}}{\text{N}^+} = \frac{\text{O}}{\text{O}^+}
\end{equation}
Under the same assumption, we calculate the abundance ratio N/O as
\begin{equation}
\frac{\text{N}}{\text{O}} = \frac{\text{N}^+}{\text{O}^+}
\end{equation}
We find that 19985 and 20062 have log(N/O)=$-0.75\pm0.12$ and $-0.89\pm0.19$, respectively, consistent with the
 solar value of log(N/O$)_\odot=-0.86$ \citep{asp09}.

\sph\ is calculated using \temop\ and [S\ii]$\lambda\lambda$6718,6733/H$\alpha$.
Since 19985 and 20062 have comparable amounts of O in O$^+$ (13.6~eV) and O$^{2+}$ (35.1~ev),
 there is likely a significant amount of S in S$^{2+}$ (23.3~eV) and S$^{3+}$ (34.8~eV) in addition to S$^+$.
%In low-excitation targets, the assumption that S/O=S$^+$/O$^+$ has been found to provide a robust correction,
% but we find 
Reliable ionization correction factor prescriptions for S are available when S$^+$ and S$^{2+}$ are observed
 \citep[e.g.,][]{thu95,dor16}.
However, S$^{2+}$ requires observation of one of the [S\iii]$\lambda\lambda$9071,9533
 lines which fall at $\approx3$~$\mu$m
 at $z=2.18$ and thus are only accessible from space.
Accordingly, we cannot derive a total S/H abundance with the current MOSFIRE data set.

\section{Discussion}\label{sec:discussion}

\subsection{Implications for metallicity calibrations at high redshift}

\begin{figure*}
 \includegraphics[width=\textwidth]{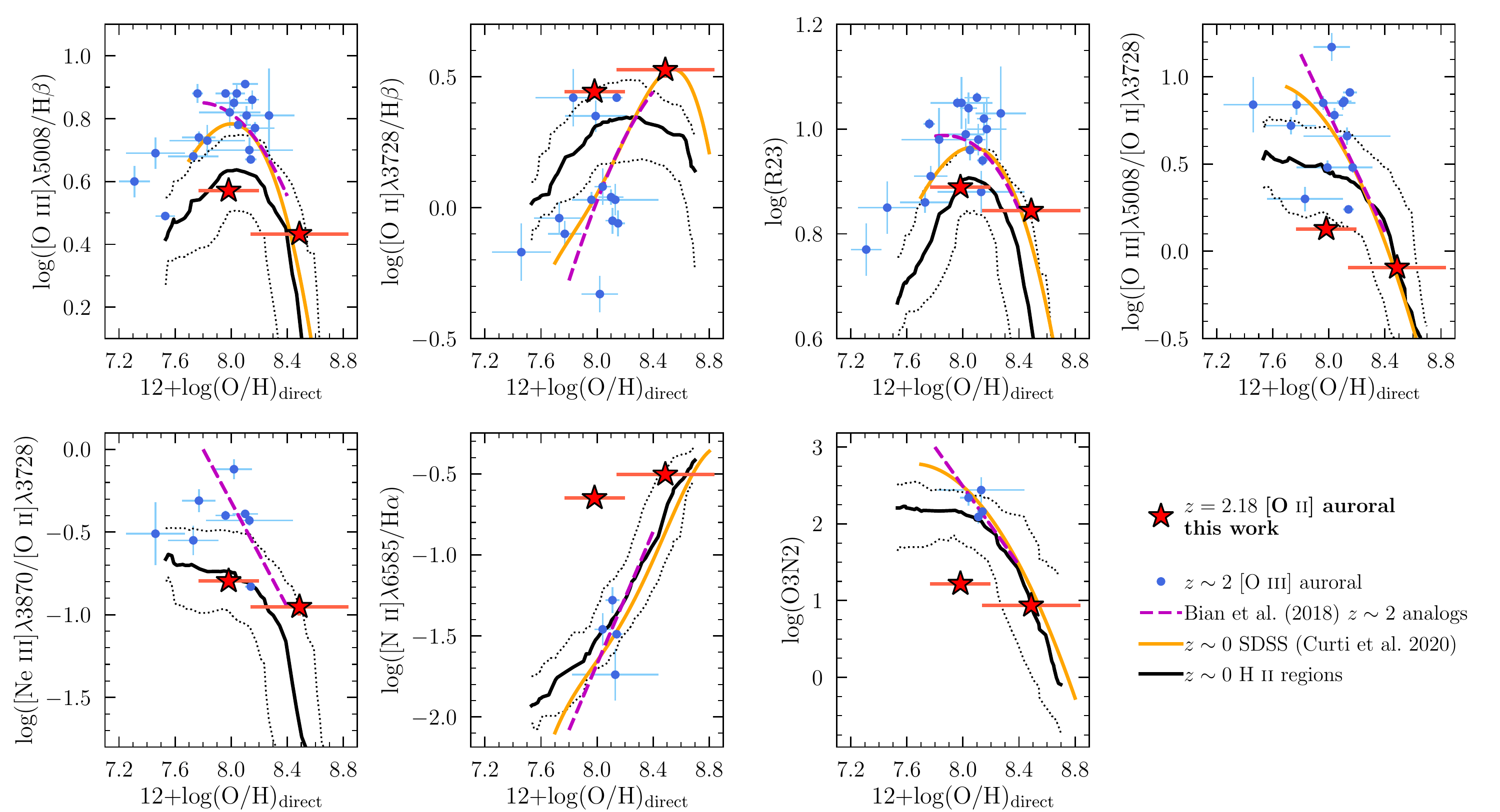}
 \centering
 \caption{
Strong-line ratios vs.\ direct-method oxygen abundance.
The two $z=2.18$ targets of this analysis, with metallicities based on [O\ii] auroral lines, are displayed as red stars.
Blue circles denote $z>1$ sources drawn from the literature with direct-method metallicities based on [O\iii] auroral lines
 \citep{san20,gbu19}.
The black solid and dotted lines show the median and $1\sigma$ scatter of the distribution of $z=0$ \hii\ regions from
 the sample presented in \citet{san17} (see also \citealt{pil16}).
The solid orange line shows the calibration set derived by \citet{cur20b} based on stacked spectra of $z\sim0$ star-forming
 galaxies from SDSS.
The dashed purple line presents the high-redshift analog calibrations of \citet{bia18}, constructed by stacking spectra
 of a sample of local galaxies selected to lie in the same region of the [O\iii]/H$\beta$ vs.\ [N\ii]/H$\alpha$ ``BPT''
 diagram as $z\sim2$ star-forming galaxies.
}\label{fig:cal}
\end{figure*}

Determining the form of strong-line metallicity calibrations at high redshift is one of
 the most pressing matters for galaxy evolution studies in the next decade.
Such relations are required to take full advantage of data from large spectroscopic galaxy surveys,
 including existing spectra for thousands of galaxies at $z\sim1-4$
 \citep[e.g.,][]{ste14,kri15,mom16,kas19a},
 and upcoming data from {\it JWST} at $z>4$ and reaching into the epoch of reionization.
In Figure~\ref{fig:cal}, we show a range of strong-line ratio metallicity indicators plotted
 against direct-method metallicity for COSMOS 19985 and 20062 (red stars).
For comparison, we include the compilation of \citet{san20} of 18 galaxies at $z\sim2.2$ with
 direct-method metallicities based O$^{2+}$ measurements (from either [O\iii]$\lambda$4364 or O\iii]$\lambda\lambda$1661,1666),
 as well as a $z=2.59$ dwarf galaxy with detected [O\iii]$\lambda$4364 from \citet{gbu19} (blue points).
We also show the calibrations of \citet{cur20b} derived from stacked $z\sim0$ SDSS spectra and those
 of \citet{bia18} derived from stacked spectra of $z\sim2$ analogs selected from SDSS.
The running median and scatter of the local \hii\ region sample presented in \citet{san20} is also displayed.

The [O\iii]-based sample drawn from the literature ubiquitously displays a high degree of excitation and ionization
 ([log([O\iii]/H$\beta)\gtrsim0.7$, log([O\iii]/[O\ii]$)\gtrsim0.5$, log(R23$)\gtrsim0.9$)
 and low metallicity (12+log(O/H$)\lesssim8.1$).
In contrast, 19985 and 20062 have lower levels of excitation and ionization, and 20062 has a higher metallicity.
This comparison demonstrates the potential of auroral [O\ii] to provide direct-method abundances in
 regions of parameter space where auroral [O\iii] is weaker and thus harder to detect.
In local \hii\ regions, it has long been known that [O\iii]$\lambda$4364 becomes difficult to detect
 at high metallicities due to the combined effect of low O$^{2+}$/O and cool $T_e$
 \citep[e.g.,][]{ber15,cro15,cro16}.
In contrast, \oiia\ remains relatively strong over a wide range in metallicity, including at low metallicities despite
 the fact that it traces a low-ionization species.
For example, in the sample of \citet{izo06} selected from SDSS based on [O\iii]$\lambda$4364 detection,
 which spans 12+log(O/H$)\sim7.7-8.5$,
 nearly every galaxy also has a detection of \oiia.
Constructing a representative sample of high-redshift galaxies with direct-method abundances spanning a wide range of
 metallicities will thus require a sample based on a mixture of high- and low-ionization auroral emission lines.
It is clear that relying on the commonly-used [O\iii]$\lambda$4364 alone will likely result in a sample that is biased toward
 high excitation and may fail to span a wide dynamic range in O/H.
Indeed, in order to trace the actual shape of the calibration between strong-line ratios and metallicities,
 a significantly wider range in 12+log(O/H) must be probed than is present in the current [O\iii]-based sample.

Considering the $z\sim2$ [O\iii]- and [O\ii]-auroral samples as a whole,
 we begin to resolve the qualitative shape of
 the relations between strong-line ratios and direct-method O/H for the first time at high redshift.
We find that [O\iii]/H$\beta$ and R23 are double valued with a turnover region around 12+log(O/H$)\sim8.0$.
[O\ii]/H$\beta$ increases with increasing metallicity up to near-solar O/H.
[O\iii]/[O\ii] and [Ne\iii]/[O\ii] both decrease with increasing metallicity, though these ratios
 display a large scatter at fixed O/H.
[N\ii]/H$\alpha$ increases and O3N2 decreases with increasing O/H.
All of these trends are in qualitative agreement with the shape of empirical and theoretical
 metallicity calibrations constructed for use in the low-redshift universe
 \citep[e.g.,][]{mcg91,kew02,pet04,mai08,pil16,cur17,cur20b,kew19}.
However, as noted in \citet{san20}, evolution in normalization at fixed O/H appears to be present on average for
 some line ratios, as evidenced by the high values of R23 and [O\iii]/H$\beta$ that local samples and calibrations
 fail to reach.

Despite the significant observational investment that has enabled this combined high-redshift auroral-line sample,
 both the sample size and precision of individual metallicity determinations are clearly too small to draw any
 quantitative conclusions about the form of high-redshift calibrations and robustly constrain their evolution
 with respect to local relations.
Even when pushing to the limits of what is currently feasible with 8$-$10~m ground-based telescopes,
 auroral lines can only be detected for the brightest line emitters at $z\sim1-3$ and in many cases require
 rare strong gravitational lensing of targets in particular redshift intervals.
Continued progress in this area clearly requires new facilities with improved capabilities.

\subsection{Prospects for direct-method abundances with {\it JWST}}

With the recent advent of {\it JWST} science operations, galaxy evolution studies will be revolutionized
 by the unrivaled infrared spectroscopy capabilities the telescope offers.
Indeed, the promise of {\it JWST}/NIRSpec to transform high-redshift direct-method metallicity studies is already
 being demonstrated with the recent detection of [O\iii]$\lambda$4363 in three lensed $z>7$ galaxies \citep{sch22,car22}.
%Much of the improvement {\it JWST} will offer comes by removing the difficulties that observing through
% the atmosphere introduces to near-infrared observations.
One of the key improvements offered by {\it JWST} for near-infrared spectroscopic observations is the removal of
 the challenges associated with observing through the Earth's atmosphere.
These include the bright and strongly wavelength-dependent background noise, and the significant wavelength
 gaps in atmospheric transmission. 
Given the faintness of auroral emission lines, an increase in sensitivity is the key requirement
 to improve upon the current high-redshift direct-method metallicity sample.
{\it JWST}/NIRSpec will yield a factor of several gain in sensitivity relative to Keck/MOSFIRE for emission-line
 studies, with the largest improvements at redder wavelengths.

As an example, for the targets of this analysis, measuring the \oiia\ lines at 2.3~$\mu$m with fluxes of
 $\sim8\times10^{-18}$~\cgsflux\ at $3-4\sigma$ required  5.5~hours of integration with Keck/MOSFIRE.
The {\it JWST} exposure time calculator suggests that NIRSpec microshutter array spectroscopy
 would enable detection of these lines at $5\sigma$
 in an on-source integration time of only $\sim$10~minutes with the G235M/F170LP setting.
Likewise, unlensed galaxies in the current [O\iii] auroral sample have [O\iii]$\lambda$4364 fluxes
 of $\sim5\times10^{-18}$~\cgsflux\ \citep{san20}.
At $z=2.2$ and 1.4~$\mu$m, this line would be detected by {\it JWST}/NIRSpec at $5\sigma$ in $\sim$30~minutes on source
 in the G140M/F100LP setting.
This exercise demonstrates how quickly {\it JWST}/NIRSpec's capabilities could reproduce the existing $z\sim2$ auroral-line
 sample, representing dozens of hours of 8$-$10~m telescope time, with a higher average S/N.
We note, however, that both the targets of the current analysis and those in the auroral [O\iii] literature sample
 have emission line fluxes considerably brighter than what is typical for $z\sim2$ main-sequence galaxies.
Consequently, deeper integrations with {\it JWST}/NIRSpec of up to several hours will enable the detection of auroral emission lines
 in more representative $z\sim2$ galaxies, moving beyond the highly-biased sample that is available from ground-based observations.
Indeed, current 8$-$10~m class ground-based facilities are simply not capable of establishing direct-method metallicity
 calibrations for a large and representative sample of star-forming galaxies at $z\sim2$.

\begin{figure}
 \includegraphics[width=\columnwidth]{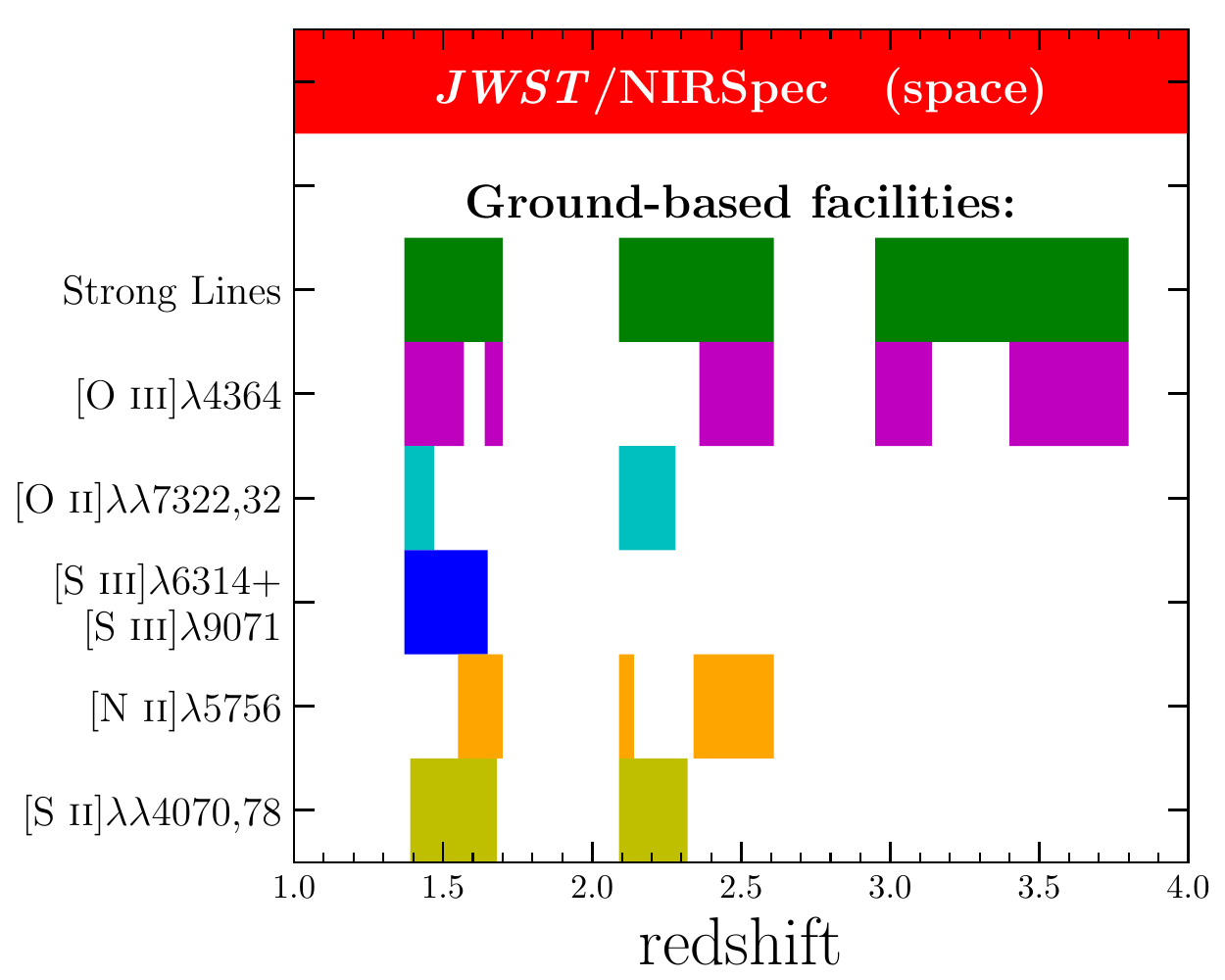}
 \centering
 \caption{
The redshift ranges across which the strong rest-optical emission lines
 ([O\ii]$\lambda$3728, H$\beta$, [O\iii]$\lambda$5008, H$\alpha$, [N\ii]$\lambda$6585, [S\ii]$\lambda\lambda$6718,6733)
 and various auroral emission lines can be observed simultaneously.
The green shaded regions show where the full suite of strong lines can be accessed from the ground in near-infrared windows
 of atmospheric transmission.
The shaded regions in each row below that present the range at which auroral lines of
 various ionic species can be observed alongside the strong lines.
Thanks to its continuous $0.7-5.0$~$\mu$m wavelength coverage, {\it JWST}/NIRSpec can access all of these lines
 across the full $z=1-4$ redshift range.
}\label{fig:zrange}
\end{figure}

In addition to an increase in sensitivity, the move to space with {\it JWST} provides a critical improvement
 for auroral line studies by removing the restriction of observing only at wavelengths
 inside the near-infrared windows of high atmospheric transmission.
Deriving direct-method metallicities requires simultaneous coverage of at least one auroral line
 (e.g., [O\iii]$\lambda$4364, \oiia) and a suite of strong rest-optical emission lines
 (i.e., H$\alpha$, H$\beta$, [O\iii]$\lambda$5008, [O\ii]$\lambda\lambda$3727,3730, [S\ii]$\lambda\lambda$6718,6733).
Fitting this combination of lines inside ground-based near-infrared bands drastically restricts the range of allowed redshifts,
 resulting in a vastly reduced number of potential targets on top of the requirement of extraordinarily bright line emission
 (indeed, 19985 and 20062 are among the brightest $z\sim2$ line emitters in the CANDELS fields).
Figure~\ref{fig:zrange} displays the redshift ranges in which the rest-optical strong lines and auroral lines of
 [O\iii], [O\ii], [S\iii], [N\ii], or [S\ii] are accessible in the ground-based near-infrared bands.
It is clear that the redshift range of possible targets is severely limited from the ground for this science case.
There is a particularly drastic limitation for all auroral lines except [O\iii]$\lambda$4364,
 partially explaining why only auroral lines of [O\iii] had been detected at $z>1$ prior to this work.
Furthermore, the [O\iii] and [O\ii] auroral emission lines can only be simulatneously accessed from the ground in the
 highly restrictive redshift range of $z=1.37-1.47$.

% hiiregiondata/talk_plots.py auroral_redshift_ranges.pdf

In contrast, {\it JWST}/NIRSpec's continuous high-sensitivity spectral coverage across $0.7-5$~$\mu$m
 enables simultaneous measurements of all rest-optical auroral lines along with the strong lines.
This capability has two major advantages for direct-method metallicity studies.
First, the number of potential auroral-line targets within the range of {\it JWST}/NIRSpec's increased sensitivity
 is greatly increased, making efficient multiplexing possible.
Such a gain in multiplexing provides significant improvement over ground-based observations with, e.g.,
 Keck/MOSFIRE for which pointings with even a handful of
 bright enough targets within the required redshift range are exceedingly rare.
Second, the continuous wavelength coverage afforded by {\it JWST} provides the ability to directly constrain
 the electron temperature in both the low- and high-ionization zones simultaneously by detecting auroral lines
 of low (e.g., [O\ii], [N\ii], [S\ii]) and high (e.g., [O\iii], [S\iii]) ionization energy species in individual targets.
This step is required to bring high-redshift chemical abundance studies on par with the level of precision
 regularly reached by local studies, and is only possible within extremely limited redshift ranges from the ground.
In this work, as in all past analyses of high-redshift auroral detections, $T_e$ is only measured in one
 zone while $T_e$ of the other zone is inferred through empirical or theoretical relations between low- and
 high-ionization temperatures \citep[e.g.,][]{cam86,izo06}.
This approach introduces systematic uncertainties associated with both the unknown form of $T_e(\text{high})-T_e(\text{low})$
 relations at high redshift and the scatter in these relations,
 the latter of which is significant at $z\sim0$ \citep[$\sim$1,000~K,][]{cro16,ber20,rog21}.
While such systematic effects are not a dominant source of uncertainty in this analysis since the \oiia\ lines are detected at
 low significance ($3-4\sigma$), they will become important as {\it JWST} enables high-S/N auroral line measurements at $z>1$.
%Establishing the form of $T_e(\text{high})-T_e(\text{low})$ 

To demonstrate the feasibility of detecting multiple auroral lines for bright targets with {\it JWST}/NIRSpec,
 we combined \temop\ and \temotp\ reported in Table~\ref{tab:temden} with the strong-line fluxes and \ebvgas\ to
 estimate the observed (reddened) fluxes of auroral lines that are inaccessible from the ground for COSMOS 19985
 due to its redshift, adopting a ratio of [S\iii]$\lambda\lambda$9071,9533/[S\ii]$\lambda\lambda$6718,6733=1.0 \citep{san20b}
 since strong [S\iii] is also unreachable from the ground at this redshift.
In units of $10^{-18}$~\cgsflux, we predict that $F_{\text{obs}}$([O\iii]$\lambda$4364)=5.4,
 $F_{\text{obs}}$([N\ii]$\lambda$5755)=2.3,
 $F_{\text{obs}}$([S\iii]$\lambda$6312)=1.8,
 and $F_{\text{obs}}$([S\ii]$\lambda$4076)=0.5.
Thus, the [O\iii], [N\ii], [S\iii], and [O\ii] auroral lines would be detectable at $\gtrsim5\sigma$ with {\it JWST}/NIRSpec MSA
 integrations of $\sim30$~minutes in G140M/F100LP and $\sim1$~hour in G235M/F170LP for targets like 19985,
 noting that this source has
 atypically bright emission lines and that integrations of several hours would be required to obtain similar results for
 galaxies closer to the star-forming main sequence.
This exercise also demonstrates that the [O\iii] and [O\ii] auroral lines are typically the brightest $T_e$ diagnostics
 at low and moderate metallicities most relevant for high-redshift studies ($\lesssim0.5$~$Z_\odot$)
 and present a technically feasible avenue to simultaneous low- and high-ionization constraints.
Indeed, for 19985, we would have detected [O\iii]$\lambda$4364 at $\sim5\sigma$ based on the depth of our $H$-band MOSFIRE observations,
 but this line fell in the atmospheric gap between the $H$ and $J$ bands.

In summary, the increased sensitivity and wavelength coverage of {\it JWST}/NIRSpec will yield more than an order of magnitude
 increase in the efficiency of detecting auroral emission lines at $z\sim1-4$ relative to the ability of
 ground-based 8$-$10~m class telescopes.
This performance promises to usher in an era of precision chemical evolution studies of high-redshift galaxies
 early in the mission lifetime of {\it JWST}.

\subsection{Minimizing systematic uncertainties on auroral [O\ii]-based metallicities}

Basing direct-method metallicities on \oiia\ presents some challenges that may systematically impact the outcome,
 but can be addressed with upcoming observations.
First, auroral \oiia\ and strong \oiis\ are widely separated in wavelength such that the strong-to-auroral ratio
 and derived \temop\ are sensitive to the reddening correction.
For example, assuming the \citet{car89} extinction curve\footnote{\citet{red20}
 found that the nebular attenuation curve in a sample of $z\sim2$ star-forming galaxies from the MOSDEF
 survey is similar to the \citet{car89} Milky Way extinction law on average.},
 we find that changing \ebvgas\ by 0.1~mag results in a change
 to \oiia/\oiis\ of 30\%, which would strongly affect the derived \temop\ and O$^+$/H$^+$.
We found that \ebvgas\ derived from higher-order Balmer lines are generally consistent
 with our fiducial value based on H$\alpha$/H$\beta$, suggesting a robust reddening correction for \oiis.
Nevertheless, significant uncertainty about the high-redshift nebular extinction curve and its variation among
 individual galaxies remains.
Estimates of dust reddening and determinations of the nebular dust law will soon be significantly improved via
 {\it JWST}/NIRSpec's long-wavelength coverage that provides access to relatively unreddeneed Paschen series lines \citep{red20}.

Temperature determinations based on \oiia\ emission can also be biased by dielectric recombination into the
 upper level that produces these transitions, leading to an overestimate of \temop\ \citep{rub86,liu00}.
However, this effect is strongest at low temperatures ($\sim5,000-8,000$~K) and high metallicities ($\gtrsim Z_\odot$),
 as well as high densities ($n_e>1,000$~cm$^{-3}$).
This effect is thus not expected to be important at high redshift where galaxies are typically
 relatively metal-poor and have $n_e\sim200-300$~cm$^{-3}$ \citep[e.g.,][]{shim15,san16a}.
Using the O\ii\ recombination coefficients from \citet{sto17}, we find that the emissivity of the
 \oiia\ lines resulting from recombination are $\gtrsim100\times$ lower than the emissivity due to
 collisional excitation at the \temop\ and $n_e$ derived for these targets.
Accordingly, recombination effects have a negligible impact on our results.

\subsection{N/O and $\alpha$/Fe abundance patterns}

\begin{figure}
 \includegraphics[width=\columnwidth]{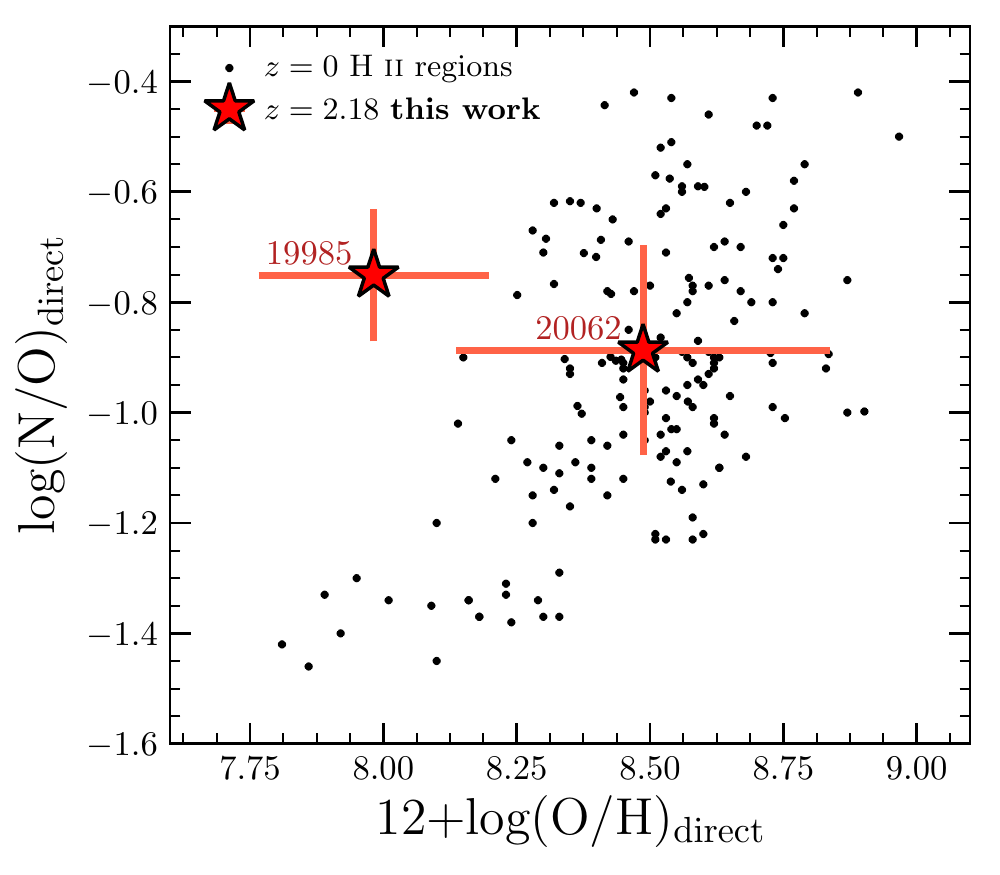}
 \centering
 \caption{
N/O vs.\ O/H for the $z=2.18$ galaxies analyzed in this work (red stars)
 and local \hii\ regions from the CHAOS survey \citep[black points;][]{ber20}.
All abundances are determined using the direct method.
19985 displays a significant offset from the $z=0$ relation.
}\label{fig:no}
\end{figure}

Chemical abundance patterns of non-$\alpha$ elements (e.g., N, C, Fe) relative to O can provide
 strong constraints on the formation histories of
 high-redshift systems \citep[e.g.,][]{ste16,ber16,ber19,top20a,top20b,cul21,str22}.
%In addition to the bulk gas-phase metallicity traced by O, the most abundant metal, chemical abundance patterns
% that include non-$\alpha$ elements (e.g., N, C, Fe) can provide strong constraints on the formation histories of
% high-redshift systems \citep[e.g.,][]{ste16,ber16,ber19,top20a,top20b,cul21,str22}.
We have determined N/H and N/O using the direct-method for the first time at high redshift.
Figure~\ref{fig:no} shows N/O vs.\ O/H for 19985 and 20062, along with a sample of local \hii\ regions
 from the CHAOS survey \citep{ber20}.
It is difficult to draw any conclusions about 20062 due to the large uncertainty on its O/H, where
 the $1\sigma$ bounds on O/H span the entire range of the $z=0$ sample at fixed N/O.
19985, on the other hand, presents evidence for elevated N/O at fixed O/H,
 being $2.5\sigma$ inconsistent in O/H with the mean $z=0$ relation at fixed N/O
 or $4\sigma$ inconsistent in N/O at fixed O/H.
The anomalously high N/O of 19985 explains why this galaxy is an outlier in [N\ii]/H$\alpha$
 and O3N2 vs.\ O/H (Fig.~\ref{fig:cal}), and highlights the potential pitfall of N-based metallicity indicators.

It is expected that rapidly forming galaxies will have low N/O abundance ratios because N enrichment,
 predominantly originating from intermediate-mass stars, occurs on longer timescales than that of $\alpha$
 elements that predominantly come from core-collapse supernovae.
Furthermore, low-metallicity systems (12+log(O/H$)\lesssim8.2$) typically lie on the ``primary nucleosynthesis''
 plateau at log(N/O$)\approx-1.4$ \citep[e.g.,][]{ber19}, where the increased presence of O in higher metallicity systems catalyzes
 CNO reactions resulting in an incresed output of N relative to O with increasing metallicity.
Given the extremely high specific SFR (sSFR=25~Gyr$^{-1}$) and low O/H (12+log(O/H)=7.98) of 19985,
 it is thus surprising that this galaxy presents near-solar N/O.

An overabundance of Wolf-Rayet stars could account for an excess of N produced by young massive stars \citep{mas14},
 though this scenario requires significant changes to either the upper end of the initial mass function or
 to stellar evolutionary processes.
Another possibility is that 19985 recently experienced a strong gas accretion event (possibly through a gas-rich merger
 of a lower-metallicity companion).
If a galaxy quickly accretes gas with a significantly lower metallicity than its preceding ISM metallicity,
 ISM metals will be diluted such that O/H decreases while heavy element abundance ratios including N/O
 will remain relatively unchanged \citep{kop05}.
This interpretation is consistent with the presence of a strong starburst (fueled by the accreted gas) in 19985,
 which lies an order of magnitude above the mean $z\sim2$ star-forming main sequence \citep{spe14},
 and with its large gas fraction reported by \citet{san22} based on CO ($M_\text{gas}/M_*\sim10$).
The best-fit SED model also suggests a sharply rising star-formation history
 (i.e., $t_\text{age}\ll\tau$; Table~\ref{tab:properties}).
While a similar N/O enhancement would be expected in 20062 given its comparably high sSFR,
 the large uncertainty on O/H (0.35~dex at $1\sigma$) prevents any useful constraints for this source.
 
%Recent gas accretion presents another possible route to elevated N/O at fixed O/H.
%Accounting for the large offset of 19985 from the mean $z=0$ N/O vs.\ O/H relation would require the accretion
% of a large mass of gas, possibly through a gas-rich merger with a low-metallicity dwarf.
%This recent accretion event would drive a strong burst of star formation,
% providing an explanation for the highly elevated sSFR in 19985 that lies
% an order of magnitude above the $z\sim2$ star-forming main sequence \citep{spe14}.
%However, the starburst would need to be recent enough that the resulting increase in core-collapse
% supernova rates has not yet enriched the ISM with O and other $\alpha$ elements that would dilute
% N/O back toward the mean N/O vs.\ O/H relation.
%The best-fit SED modeling parameters for 19985 indicate a very young age (30~Myr) and large $\tau$ (10~Gyr),
% indicating a strongly rising star-formation history consistent with this interpretaion.
%Stellar population modeling of the rest-frame ultraviolet continuum indicate a similarly young age of $\sim10$~Myr
% \citep{top20b}.

\citet{top20b} recently reported constraints on the Fe/H of young stars in 19985 and 20062 based on
 modeling the rest-UV continuum, finding that both systems host low-metallicity massive stars
 with $Z_*=0.001=0.07~Z_\odot$ (i.e., 12+log(Fe/H)=6.35).
Under the assumption that the chemical composition of young stars is identical to the instantaneous gas-phase ISM
 composition, comparing Fe/H from \citet{top20b} to direct-method O/H from this work provides a measure of the
 $\alpha$/Fe ratio in these systems.
We find $\alpha$/Fe=$2.9\pm1.5\times\alpha/\text{Fe}_{\odot}$ and $8.6_{-4.3}^{+10.0}\times\alpha/\text{Fe}_{\odot}$
 for 19985 and 20062, respectively, consistent with  $\alpha$-enhancement relative to the solar abundace pattern
 for both galaxies.\footnote{If the $T_e$-based metallicities are converted to the O recombination line abundance scale,
 as has been suggested by some recent studies \citep[e.g.,][]{bla15,ste16,san20}, the O/H and inferred $\alpha$/Fe values
 would increase by a factor of 1.7}
This result agrees with other recent studies
 that find typical star-forming galaxies at $z\sim2-3$ are $\alpha$-enhanced by factors of $\sim2-5$ relative to
 the solar abundance pattern \citep{ste16,str18,str22,cul21,san20,top20a,top20b,red22}.
Such $\alpha$-enhancement occurs due to their rapid formation timescales that favor significant $\alpha$ enrichment
 from prompt core-collapse supernovae but have not yet been significantly enriched in Fe-peak elements by
 Type~Ia supernovae that occur on longer timescales.
This $\alpha$-enhancement (or Fe-deficit) has important implications for the hardness of the ionizing spectrum in \hii\ regions,
 a crucial consideration for interpreting emission-line ratios,
 since Fe-peak line blanketing is a major factor governing the ionizing photon output of massive stars.

These results demonstrate the power of chemical abundance patterns to shed light on galaxy formation histories.
Novel measurements with {\it JWST} will provide the necessary improvement on the precision of direct abundancess
 to transform such analyses from a qualitative to quantitative regime at high redshift.
The wide and continuous spectral coverage of {\it JWST}/NIRSpec, alongside the ability to simultaneously constrain
 $T_e$ in both the low- and high-ionization zones, will enable the determination of multi-element chemical abundance
 patterns at $z>1$ approaching the level of detail present in local studies \citep[e.g.,][]{izo06}.

\subsection{Is the line emission powered by AGN?}\label{sec:agn}

It is of interest to consider whether the line emission in these galaxies is powered
 by the accreting black hole of an AGN instead of photoionization
 by massive stars in \hii\ regions, as has been implicitly assumed in this analysis.
Both targets are offset from the mean $z\sim0$ sequence of star-forming galaxies in
 the [O\iii]/H$\beta$ vs.\ [N\ii]/H$\alpha$ ``BPT'' diagram \citep{bpt81},
 falling in the ``composite'' region between local star-forming galaxies and AGN \citep{kew01,kew06,kau03}
 but squarely within the distribution of typical $z\sim2$ star-forming galaxies \citep[e.g,][]{ste14,sha15,str17,run22}.
However, these targets are both within the pure star-forming region of the [O\iii]/H$\beta$ vs.\ [S\ii]/H$\alpha$
 and [O\n]/H$\alpha$ diagnostic diagrams, where local AGNs are observed to have log([S\ii]/H$\alpha)>-0.5$
 and log([O\n]/H$\alpha)>-1.4$ at similar [O\iii]/H$\beta$ as 19985 and 20062 \citep{vei87,kew06}.
The [O\iii]/H$\beta$ vs.\ \oiia/H$\alpha$ diagram has also been proposed as a diagnostic to distinguish between
 star-formation and AGN-ionized sources \citep{ost92}.
We measure dust-corrected log(\oiia/H$\alpha)=-1.58\pm0.08$ and $-1.69\pm0.14$ for 19985 and 20052, respectively.
These values again place our targets within the distribution of star-forming galaxies and \hii\ regions.
Low-redshift AGN and composite objects have log(He\ii~$\lambda$4686/H$\beta)>-1.4$ \citep[e.g.,][]{shi12}.
While He\ii~$\lambda$4686 is not detected in either spectrum, we place stringent $3\sigma$ upper limits of
 log(He\ii~$\lambda$4686/H$\beta)<-1.64$ and $<-1.38$ for 19985 and 20062, respectively.

% other line ratios:
% 19985: log(oii7325/ha) = -1.58 +/- 0.08
% log(ariii7135/oiii5007) = -1.91 +/- 0.08
% log(heii4686/hb) 3sigma < -1.64
% 20062: log(oii7325/ha) = -1.69 +/- 0.14
% log(ariii7135/oiii5007) = -1.79 +/- 0.09
% log(heii4686/hb) 3sigma < -1.38
% stasinska 2006 calibration: 12 + log O/H = 8.91 + 0.34x + 0.27x^2 + 0.20x^3
% x = log(ariii7315/oiii5007)
% ar3o3 metallicities: 7.85pm0.15, 8.01pm0.12 or 19985, 20062

Furthermore, 19985 and 20062 are not detected in X-rays in the Chandra observations of the COSMOS field,
 nor do they display Spitzer infrared colors indicative of AGN-heated dust \citep{coi15,aza17}.
Both targets additionally have deep Keck/LRIS spectra covering the rest-frame ultraviolet
 over 1000$-$2300~\AA\ \citep{top20a,top20b}, in which no high-ionization emission lines
 typical of AGN (e.g., N~\textsc{v}, He\ii, C\iv) are detected.
In summary, a wide range of diagnostics including X-ray, infrared, and rest-frame optical and ultraviolet data
 strongly suggest that AGN ionization is not a significant contributor to the total line emission in these galaxies
 and thus does not affect our results.

\subsection{The presence of outflows and broad emission}

Strongly concentrated star-formation has been found to be associated with
 strong and efficient gas outflows \citep[e.g.,][]{ste10,new12,hec15,davi19,for19,wel22}.
The extremely high SFR surface densities (\sigsfr=18.5 and 19.8~$\msun~\text{yr}^{-1}~\text{kpc}^{-2}$)
 of 19985 and 20062 suggest that the starbursts they are currently experiencing are likely driving powerful outflows.
Indeed, both galaxies are observed to have blueshifted and asymmetric rest-UV absorption lines with extended blue wings \citep{wel22}.
Furthermore, \citet{leu17} fit rest-optical emission lines in the MOSDEF spectra of 19985 and 20062
 with double-Gaussian profiles including a broad and narrow
 component,\footnote{\citet{leu17} analyzed these targets as AGN based on a BPT diagram selection. With additional data,
 we now find strong evidence against the presence of AGN in these galaxies (Sec.~\ref{sec:agn}).}
 detecting broad components in both galaxies with FWHM$\sim$500~km~s$^{-1}$ and offset from the systemic redshift
 by $\sim-50$~km~s$^{-1}$.
In the strong rest-optical lines ([O\ii], H$\beta$, [O\iii], H$\alpha$, and [N\ii]), we observe wide asymmetric emission
 with a more prominent blue wing that the single-Gaussian profiles fail to fit, though this ``missed" emission is
 $\lesssim5\%$ of the single-Gaussian line flux.

It is thus clear that strong star formation is driving significant outflows in both galaxies,
 detected in both neutral and ionized phases.
We will present a more detailed analysis of the outflow properties of these targets in future work.
What concerns the current analysis is whether the presence of such outflows biases the inference of
 physical properties and chemical abundances from integrated emission line measurements.
The answer to this question ultimately comes down to 
 the fraction of the total line fluxes comprised by the broad component,
 and the magnitude of the difference between the physical conditions in the outflowing gas and \hii\ regions.
If the broad component contributes a small fraction of the total flux, it is unlikely to strongly
 bias properties derived from the integrated lines even if the physical conditions differ significantly.
However, if the outflowing component contributes a fraction of the total flux comparable to the narrow ISM
 component, inferences from integrated lines may be biased.
A more detailed analysis beyond the scope of this work is required to robustly address these questions.
As {\it JWST} enables the measurement of temperature and density diagnostics at higher redshifts when galaxies
 had higher SFRs and smaller radii (and thus are expected to have stronger outflows),
 it will become increasingly important to understand whether outflowing gas biases abundance determinations
 from integrated line measurements.

%Ionized outflows at $z\sim2$ have been found to have somewhat larger electron densities ($n_e\sim400$~cm$^{-3}$)
% than the narrow ISM component \citep{for19}, though these densities are still low enough that
% derived electron temperatures and abundances will not be significantly affected.
%In contrast, the inferred ISM metallicities will be strongly affected if the outflowing gas has significantly different
% electron temperatures and abundances than gas in \hii\ regions.
%\citet{cam21}

\section{Summary and Conclusions}\label{sec:summary}

In this paper, we presented ultra-deep rest-optical spectroscopy of 
 two star-forming galaxies at z=2.18 in the COSMOS field with bright rest-optical emission lines,
 representing more than 20~hours of on-source integration with Keck/MOSFIRE.
The high fidelity of the resulting spectra enabled the first detections of
 the auroral \oiia\ emission-line doublet outside of the low-redshift universe.
In turn, we used these data to obtain constraints on the
 electron temperature in the low-ionization zone of the ionized ISM and calculate the
 gas-phase O/H, N/H, and N/O abundance ratios via the direct method.
Our main results are summarized as follows:

\begin{enumerate}

\item We measured low-ionization temperatures of \temop=12,440$\pm$1680~K and 9,330$\pm$1,350~K
 and gas-phase oxygen abundances of 12+log(O/H)=$7.98$$\pm$$0.22$ and $8.49$$\pm$$0.35$
 for COSMOS 19985 and 20062, respectively.
These detections of \oiia\ at $z\sim2$ demonstrate the feasibility of using [O\ii] auroral
 lines for direct-method abundance studies of distant galaxies in the early universe.
The NIRSpec instrument on {\it JWST} would take only $\sim10$~minutes of exposure at $R\sim1,000$
 to detect \oiia\ as bright as observed in these targets.
As such,  auroral [O\ii] can be detected in the brightest high-redshift line emitters with
 only shallow integrations with {\it JWST}/NIRSpec.

\item We placed the new [O\ii]-based direct-method O/H measurements alongside a sample of
 $\sim20$ galaxies at $z>1$ with direct-method metallicities based on auroral [O\iii]
 on diagrams of strong-line ratio vs.\ 12+log(O/H)$_\text{direct}$.
We found that the [O\ii]-based targets have lower levels of ionization and excitation and lie
 at higher metallicity than the bulk of the [O\iii]-based sample.
This result suggests that obtaining a representative direct-method abundance sample, essential for producing accurate
 strong-line calibrations for use at high redshifts, will ultimately require a combination of low-
 and high-ionization auroral line measurements to populate a sufficient dynamic range in metallicity and excitation.

\item {\it JWST}/NIRSpec promises to simultaneously detect low- and high-ionization auroral lines of relatively bright
 individual targets in moderately deep exposures ($\ge1$~hour), with [O\iii] and [O\ii] being the most observationally accessible
 combination.% at most metallicities.%, while [S\iii] and [N\ii] are also feasible possibilities.
The gain in efficiency of {\it JWST}/NIRSpec over ground-based facilities for auroral-line surveys is two-fold,
 benefiting from both an increase in sensitivity and from uninterrupted wavelength coverage
 over $1-5$~$\mu$m (where ground-based facilities can only observe in segmented near-infrared atmospheric transmission windows).
These combined effects significantly increase the on-sky density of potential targets for which
 the required auroral and strong lines can be simultaneously observed, significantly improving multiplexing efficiency.

\item We investigated the abundance patterns of non-$\alpha$ elements, including N and Fe.
One galaxy, 19985, displays near-solar N/O despite having $\sim20$\% solar O/H.
This unexpected composition may indicate a recent accretion of a large mass of relatively unenriched gas.
This scenario is consistent with the apparent starburst nature of this target.
%, which lies
% an order of magnitude above the $z\sim2$ star-forming main sequence,
% has a high gas fraction of $M_\text{gas}/M_*\sim10$, and a strongly rising
% star-formation history from SED modeling.
We found evidence of super-solar $\alpha$/Fe in both objects by comparing the direct-method O/H
 to Fe/H derived from modeling the rest-UV continuum by \citet{top20b}.
This $\alpha$-enhancement suggests rapid formation timescales for these systems, consistent with
 other studies at $z\sim2-3$.
The ability of {\it JWST}/NIRSpec to measure auroral emission lines of multiple species promises
 constraints on gas-phase chemical abundance patterns approaching the level of detail
 present in studies of local \hii\ regions and star-forming galaxies.

\item The depth of the spectra analyzed here enabled a number of diagnostic tests to determine whether
 accreting supermassive black holes contribute significantly to the line emission in these sources.
We found no evidence of AGN activity based on the position of our targets in the
 [O\iii]/H$\beta$ vs.\ [N\ii]/H$\alpha$, [S\ii]/H$\alpha$, and [O\n]/H$\alpha$ ``BPT'' diagrams.
This conclusion was further confirmed by the measured \oiia/H$\alpha$ ratio and stringent upper
 limits on the He\ii~$\lambda$4686/H$\beta$ ratio, both lower than the value displayed by active galaxies.

\item We found evidence for powerful gas outflows based on broad blueshifted wings around strong emission lines,
 revealed in the deep rest-optical spectra.
%, further confirmed by the presence of broad blueshifted rest-UV absorption lines \citep{wel22}.
The presence of ionized emission from such strong outflows may bias inferences on physical conditions and
 abundances of the ISM based on integrated line emission.
 %and raises the question of whether line emission from \hii\ regions can be robustly disentangled from that of
 %outflowing gas that may have significantly different physical properties and ionization mechanisms.
This question requires more analysis and will become increasingly important at higher redshifts where
 star formation becomes more concentrated and is more likely to drive massive galaxy-scale outflows.

\end{enumerate}

This analysis has demonstrated the feasibility of detecting auroral \oiia\ to constrain the
 chemical abundances of high-redshift star-forming galaxies across a range of metallicities.
However, a significant observational investment was required to obtain these results on auroral
 [O\ii] emission for two galaxies and also assemble the
 small, biased, and low-precision sample of high-redshift auroral [O\iii] targets,
 representing many nights on 8$-$10~m ground-based facilities.
The significant cost of these efforts clearly demonstrates that
 current ground-based telescopes cannot provide the quality of near-infrared spectroscopy required
 to robustly understand the absolute metallicities of high-redshift sources.
The advent of {\it JWST}, and moderate-resolution spectroscopy with its NIRSpec instrument in particular,
 represents an unprecedented leap forward in our ability to efficiently detect temperature-sensitive
 auroral emission lines at high redshift,
 ushering in an era of precision chemical abundance studies in the early universe.

\begin{acknowledgments}
Support for this work was
 provided through the NASA Hubble Fellowship
 grant \#HST-HF2-51469.001-A awarded by
 the Space Telescope Science Institute, which is
 operated by the Association of Universities for
 Research in Astronomy, Incorporated, under NASA
contract NAS5-26555.
We also acknowledge support from NSF AAG grants
AST-1312780, 1312547, 1312764, 1313171,
2009313, and 2009085, grant AR-13907 from
the Space Telescope Science Institute, grant
NNX16AF54G from the NASA ADAP program.
We wish to recognize and acknowledge the very significant cultural role and reverence
 that the summit of Maunakea has always had within the indigenous Hawaiian community.
We are most fortunate to have the opportunity to conduct observations from this mountain.
%We finally wish to extend special thanks to those of
%Hawaiian ancestry on whose sacred mountain we
%are privileged to be guests. Without their generous
%hospitality, the work presented herein would not
%have been possible.
\end{acknowledgments}

\vspace{5mm}
\facilities{Keck:I (MOSFIRE)}

\software{\texttt{pyneb} \citep{lur15}
          }

%%%%%%%%%%%%%%%%%%%%%%%%%%%%%%%%%%%%%%%%%%%%%%%%%%



\bibliography{ms}

%%%%%%%%%%%%%%%%%%%%%%%%%%%%%%%%%%%%%%%%%%%%%%%%%%

\end{document}